\documentclass[twocolumn]{aastex631}
\newcommand{\kms}{\ifmmode{\,\hbox{km\,s}^{-1}}\else {\rm\,km\,s$^{-1}$}\fi}
\shorttitle{Streams in CDM and WDM}
\shortauthors{Carlberg, Jenkins, Frenk \& Cooper}

\begin{document}

\title{Star Stream Velocity Distributions in CDM and WDM Galactic Halos}

\author[0000-0002-7667-0081]{Raymond G. Carlberg}
\affiliation{Department of Astronomy \& Astrophysics,
University of Toronto,
Toronto, ON M5S 3H4, Canada} 
\email{raymond.carlberg@utoronto.ca}

\author[0000-0003-4389-2232]{Adrian Jenkins}
\affiliation{Institute for Computational Cosmology, Department of Physics, University of Durham, South Road, Durham, DH1 3LE, UK} 
\email{a.r.jenkins@durham.ac.uk}

\author[0000-0002-2338-716X]{Carlos S. Frenk}
\affiliation{Institute for Computational Cosmology, Department of Physics, University of Durham, South Road, Durham, DH1 3LE, UK} 
\email{C.S.Frenk@durham.ac.uk}

\author[0000-0001-8274-158X]{{Andrew}~P.~Cooper}
\affiliation{Institute of Astronomy and Department of Physics, National Tsing Hua University, 101 Kuang-Fu Rd. Sec. 2, Hsinchu 30013, Taiwan}
\email{apcooper@gapp.nthu.edu.tw}

\begin{abstract}
The dark matter subhalos orbiting in a galactic halo perturb the orbits of stars in thin stellar streams. Over time the random velocities in the streams develop non-Gaussian wings. The rate of velocity increase is approximately a random walk at a rate proportional to the number of subhalos, primarily those in the mass range $\approx 10^{6-7} M_\odot$.     The distribution of random velocities  in long, thin, streams is measured in simulated Milky Way-like halos that develop in representative WDM and CDM cosmologies.   The radial velocity distributions are well modeled as  the sum of a Gaussian and an exponential. The resulting MCMC fits find Gaussian cores of $1-2\kms$ and exponential wings that increase from 3 \kms\  for 5.5 keV WDM, 4 \kms\ for 7 keV WDM, to  6 \kms\ for a CDM halo. The observational prospects to use stream measurements to constrain the nature of galactic dark matter are discussed.
\end{abstract}

\section{INTRODUCTION}

The mass distribution of subhalos orbiting within a { Milky Way-like dark halo  formed in a CDM cosmology} is a power law with thousands of subhalos to $10^6 M_\odot$ \citep{Klypin99,Moore99,Springel08}. However if the dark matter particle has a small thermal velocity, as can occur for a neutrino-like particle with a mass in the keV range,  free-streaming \citep{BS83} reduces the primordial density perturbation power \citep{Bode01} on lengths corresponding to mass scales of roughly $10^9 M_\odot$ and below. The reduction in subhalo numbers \citep{Benson13,Angulo13,Lovell14} relative to CDM  increases as the WDM particle mass decreases. Viable WDM models must have enough subhalos to account for the known dark matter dominated dwarf galaxies in the Milky Way  \citep{Newton21,Nadler21MW}  ($\geq$ 2.0 keV, $\geq$ 6.5 keV, respectively).  Strong lensing image flux ratio modeling \citep{MS98}  finds $\geq$ 6.1 keV) \citep{Keeley24} for distant galaxies. Resolved strong gravitational lensing at milli-arcsecond scales will reveal locations and masses of lower mass subhalos, whether they have stars or not \citep{StrongLens23,Vegetti24}. Each method has observational and modeling complications with accompanying systematic errors. Ultimately having several independent measures of the subhalo numbers both locally and in more distant galaxies will lead to a confident result. 

A dark matter subhalo crossing a tidal stream of stars from a globular cluster perturbs the velocities in the encounter region.  The rate of subhalo encounters is directly proportional to the number of subhalos within the radial range of the stream orbit. The stream develops a characteristic density gap over an orbital period \citep{Ibata02,Johnston02,Carlberg13}. In principle, a measurement of the number of stream gaps as a function of gap size provides a statistical measure of the subhalo mass function \citep{Carlberg12}.  A significant complication is that cluster stars are pulled into the stream with a range of angular momenta at a mass loss rate that peaks near the progenitor cluster's orbital pericenter. That is, stellar streams are created with considerable phase space structure. The stream stars have a range of orbital periods which results in a  shear in orbital angle relative to each other along the length of the stream which confuses and blurs out the gaps \citep{Ngan15,Ngan16}.,  For example,  a stream with a typical angular momentum spread $\Delta L/L=0.02$  has and angular velocity spread $\Delta \Omega/\Omega =-0.02$ (for a flat circular velocity). After, say, 3 orbits, $T=3\cdot 2\pi/\Omega\simeq 20/\Omega$, the angular spread is $\Delta\Omega T\simeq 0.4$ or about $23^\circ$. That is, a gap is only readily discerned for an orbit or two. 

The orbital shearing of stars along a stream does not alter the subhalo induced increase of random velocities in the streams. Consequently, a stream's width and its velocity spread are a measure of the density of dark matter subhalos in a galactic halo. At any given location along a stream the velocity profile is not well mixed because the rate of subhalo stream crossings is low, with about one stream crossing per 10 kpc per 4 Gyr for the expected number of subhalos around $\approx 10^{7} M_\odot$  \citep{CA23}. However, summing the velocity profile along the whole stream provides averaging. Summing along the stream length also means that the velocity profile measurement does not require the location of the progenitor, although if present that provides additional information. A variety of methods have been used to find the currently known streams in the Milky Way \citep{Mateu23}  but having a stream which stands out above the background  introduces stream width selection effects. 

The goal of this paper is to measure, model, and compare the velocity profiles of long, thin, streams in a simulated Milky Way that develops in a CDM cosmology and in 5.5 and 7 keV warm dark matter cosmologies.  Globular clusters composed of dynamically self-consistent $1 M_\odot$ star particles are inserted into the dwarf galaxy-like subhalos present at a simulation time of 1 Gyr (relative to the Big Bang), about redshift 6, to serve as old, metal poor cluster progenitors of stellar streams. The simulations also contain an imposed growing galactic disk and bulge. At the end of the simulation long, thin streams are located and the random velocity distributions about the stream centerlines averaged.  The simulations are described in the next section. The resulting halo and subhalo properties are discussed in Section~\ref{sec_halos}. The time development of the thin streams and their morphologies are discussed in Section~\ref{sec_streams}. A semi-analytic dynamical overview of subhalo interactions with streams is presented in Section~\ref{sec_dynamics}. In Section~\ref{sec_fitting} the stream velocity profiles are modeled with a dynamically based function and a simple empirical model.  Section~\ref{sec_errors} undertakes a rudimentary error analysis of velocity profile fitting to assess the samples required to provide a dark matter constraint.

\section{Simulation Setup \label{sec_sims}}

The dark matter distribution of the simulations is set up starting with the MUSIC code \citep{MUSIC}. A 40/h Mpc box of dark matter particles  is generated in the default MUSIC flat cosmology $\Omega_m=0.276$, $\Omega_b=0.045$, $H_0= 70.3$, $\sigma_8=0.811$ and $n_s=0.961$ {The results here are more dependent on the selection of a region that has a Milky Way-like assembly history, rather than the precise details of the background cosmology.}  The redshift 50 box of particles is evolved to redshift zero using the Gadget4 code \citep{Gadget4}. The AHF halo finder code \citep{AHF1,AHF2} is run on the particle distribution { at redshift zero} to identify Milky-Way like systems, which are taken to be halos with masses { within 10\% of} $10^{12} M_\odot$  with no comparable mass halos closer than 0.5 Mpc and no major mergers over the last 5 Gyr. There usually are M31-like companions in the 0.5 Mpc distance range  { Most of the 86 candidate halos in the desired mass range are rejected because they are in groups or small clusters. After the isolation and no recent mergers criteria are applied, about half a dozen reasonable candidates remain. } The region containing several of the  candidate halos are { arbitrarily} selected to be regenerated with much higher resolution. The high resolution region that MUSIC generates at redshift 50 is trimmed to be within a sphere and randomly down-sampled by about a factor of 3.3 to give 122,235,616 dark matter particles of mass 10322 $M_\odot$ with all units now converted to physical quantities. The down-sampling introduces some noise in the initial conditions which we find increases $\sigma(M)/M$ { as measured in randomly placed spheres} about 10\%  at a mass of $10^6 M_\odot$. {The high resolution regions are run forward to redshift zero with dark matter only to ensure that the mass, isolation and merger history targets remain in place. }

Warm dark matter simulations are set up starting with the \citet{Bode01} WDM power spectrum for 5.5 and 7.0 keV thermal WDM as realized in the MUSIC code. The 5.5 keV model is chosen as being marginally compatible with dwarf galaxy numbers within 60 kpc \citep{McConnachie12} slightly below the currently allowed range \citep{Nadler21MW}.  The low level of small scale power allows matter particle noise instabilities in the filaments  which develop as the simulation is evolved.  These instabilities lead to a large population of low mass subhalos below about $10^6 M_\odot$ with numbers that depend on the mass resolution of the simulation \citep{WW_wdm07}. The spurious halos can be identified as originating in abnormally flattened initial shapes and removed from the halo counts \citep{Lovell14}. Nevertheless, the spurious halos are present in the simulation and have real dynamical effects. The relatively large dark matter softening used here, 100 pc, is comparable to the initial mean particle distribution, 130 pc, which helps to suppress the filament instability. The 7 keV setup is used as created. However in the 5.5 keV simulation the number of $\simeq 10^6 M_\odot$ subhalos is about an order magnitude above the trend from larger masses. To suppress the formation of the spurious subhalos,  a random velocity is added to each dark matter particle in the redshift 50 initial setup. A velocity drawn from a Gaussian with a 3D dispersion of 4 \kms\ is applied at redshift 50 to every particle after trying a range of values from zero to 8 \kms. { The halo mass function was used to select the random velocity of 4 \kms. Higher velocities  depress the mass function at all masses, lower values leave a strong upturn at low masses.}

Globular cluster internal two-body interactions continuously cause some stars to drift outwards \citep{Spitzer87} where time varying tidal fields of the overall potential further heat the stars \citep{BT08} and eventually pull them away from the cluster into leading and trailing streams.  The globular clusters are composed of 1 $M_\odot$ star particles softened at 2~pc with an added heating term to mimic two-body interactions at the Spitzer relaxation rate \citep{Spitzer87,Carlberg18}.  All the star and dark matter particles are integrated together with Gadget4 to ensure that the stream paths and the velocities of stars within them are captured accurately with minimal assumptions. The adopted procedure follows that complex orbits of stars in the region beyond the half mass radius  as tidal forces heat and sweep the stars out of the clusters \citep{FH00,Meiron21}. 
\begin{figure}
\begin{center}
\includegraphics[angle=0,scale=0.55,trim=20 0 0 20, clip=true]{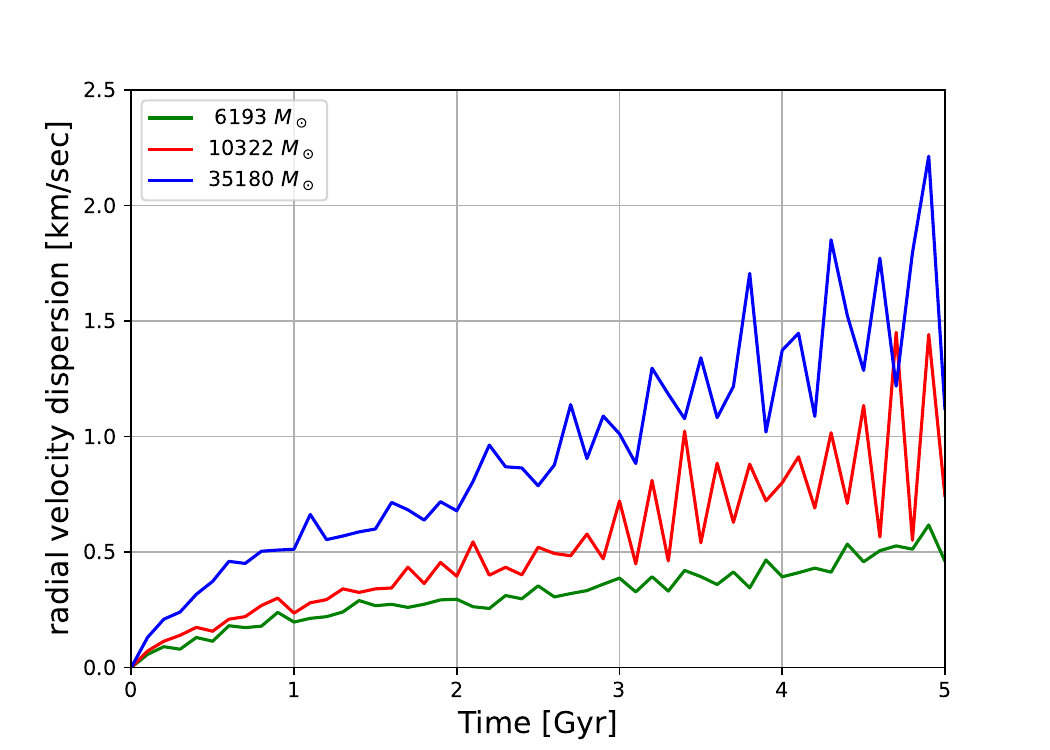}
\end{center}
\caption{
Dark matter particle heating of a ring of star particles with time for the 10322 $M_\odot$ dark matter particles in the simulations reported here, the 35180 $M_\odot$ particles of the FIRE initial conditions  \citep{Wetzel16} used in \citep{CA23}, and a lower main halo mass,  simulation with 6193 $M_\odot$ particles.
}
\label{fig_heating}
\end{figure}

The dark matter only simulations are first run from the setup time at redshift 50 to an age 1~Gyr, approximately redshift 6, to generate the dark matter distribution into which the globular clusters can be inserted. { The MUSIC code creates the initial  conditions for the CDM and WDM simulations with the same random seed so the large scale structure and the Milky Way-like halo that we identify for refinement are very nearly the same in all three simulations.}  The most massive halo at 1~Gyr in the high resolution region, $2.7\times 10^{11} M_\odot$, remains the dominant halo over the entire course of the simulations. A nuclear bulge particle with mass $5\times 10^8 M_\odot$   and softening of 0.25 kpc  is added instantaneously at 1 Gyr. {The Plummer sphere bulge  has the bulge mass of MW2014 \citep{galpy} but with a simple gravitational potential.}  The bulge has a gravitational  radius of about 0.1 kpc, so the disturbance is limited and settles down in a few dynamical times.  At 5 Gyr, shortly after a major merger { which would destroy any disk in place at that time}, a Miyamoto-Nagai disk-bulge \citep{MN75} centered on the nuclear bulge particle is inserted in the xy plane of the halo, growing linearly with time to redshift zero when it has same mass, $6.8\times 10^{10} M_\odot$,  and scale parameters, a=3, b=0.28 kpc, as a MW2014 disk \citep{galpy}. Globular clusters in the mass range $4-30\times 10^4 M_\odot$ are inserted on disk-like orbits in the subhalos more massive than $10^8 M_\odot$ following the procedure of \citet{CA23}.  The star clusters are chosen to be in the mass range which dominates the creation of current epoch streams. That is,  clusters more massive than $3\times 10^5 M_\odot$ lose mass slowly and are relatively few in number, and  star clusters below $4\times 10^4 M_\odot$ often dissolve before the end of the simulation leaving behind diffuse streams. The clusters have a half-mass radius relation approximately $r_h\simeq 5 (M/10^5 M_\odot)^{1/3}$pc. 

A complication in a mixed star and dark matter particle simulation is that the dark matter particles heat unbound star particles \citep{Chandrasekhar42}. The heating rate has an approximately logarithmic dependence on softening \citep{BT08} so is only really suppressed with increased particle resolution, that is, more, lower mass, dark matter particles. To measure the rate of star particle heating the primary halo at the end of the simulation is extracted and rerun. The dark matter particles keep their radii but are assigned random angles on a sphere to create a spherically symmetric potential. Their radial velocities are retained but the tangential velocity is randomly reoriented. A ring of star particles is inserted at 20 kpc on circular orbits.  The Miyamoto-Nagai disk is replaced with a Plummer sphere with a scale radius of 3 kpc, which causes the inner halo to be  briefly somewhat out of equilibrium. Figure~\ref{fig_heating} shows the standard deviation about the mean radial velocity for all particles in the ring with time. For the particles used here the velocity dispersion after 5 Gyr  is about 1 \kms, well below the $\sim$10 \kms\ that a typical stream particle acquires through subhalo interactions in this time.

\section{Halo and subhalo Properties \label{sec_halos}}

At redshift zero the simulations have a dominant Milky Way-like halo with $M_{200} = 9.22\times 10^{11} M_\odot$ and $r_{200}$ of 200 kpc where the 200 times critical density values are from the group finder.  Adding in the disk and bulge mass increases $r_{200}$ to 205 kpc. The WDM simulations have primary halo masses and sizes within 0.3\% of the CDM values. The gravitational mass within 50 kpc is $4.6\times 10^{11} M_\odot$, within 100 kpc is $7\times 10^{11} M_\odot$ and  $9.9\times 10^{11} M_\odot$ within 200 kpc, in a significantly triaxial halo (a=0.94, b=0.78, from AHF). The mass-radius values are close to those that \citet{Shen22} find for the Milky Way. { An exact match to the Milky Way and its globular cluster population is not required, since this paper is exploring the general properties of a stream population in a cosmologically evolving Milky Way-like potential. Matching individual observed streams in an evolving potential and subhalo population remains a research challenge.}

\begin{figure}
\begin{center}
\includegraphics[angle=0,scale=0.39,trim=40 20 20 60, clip=true]{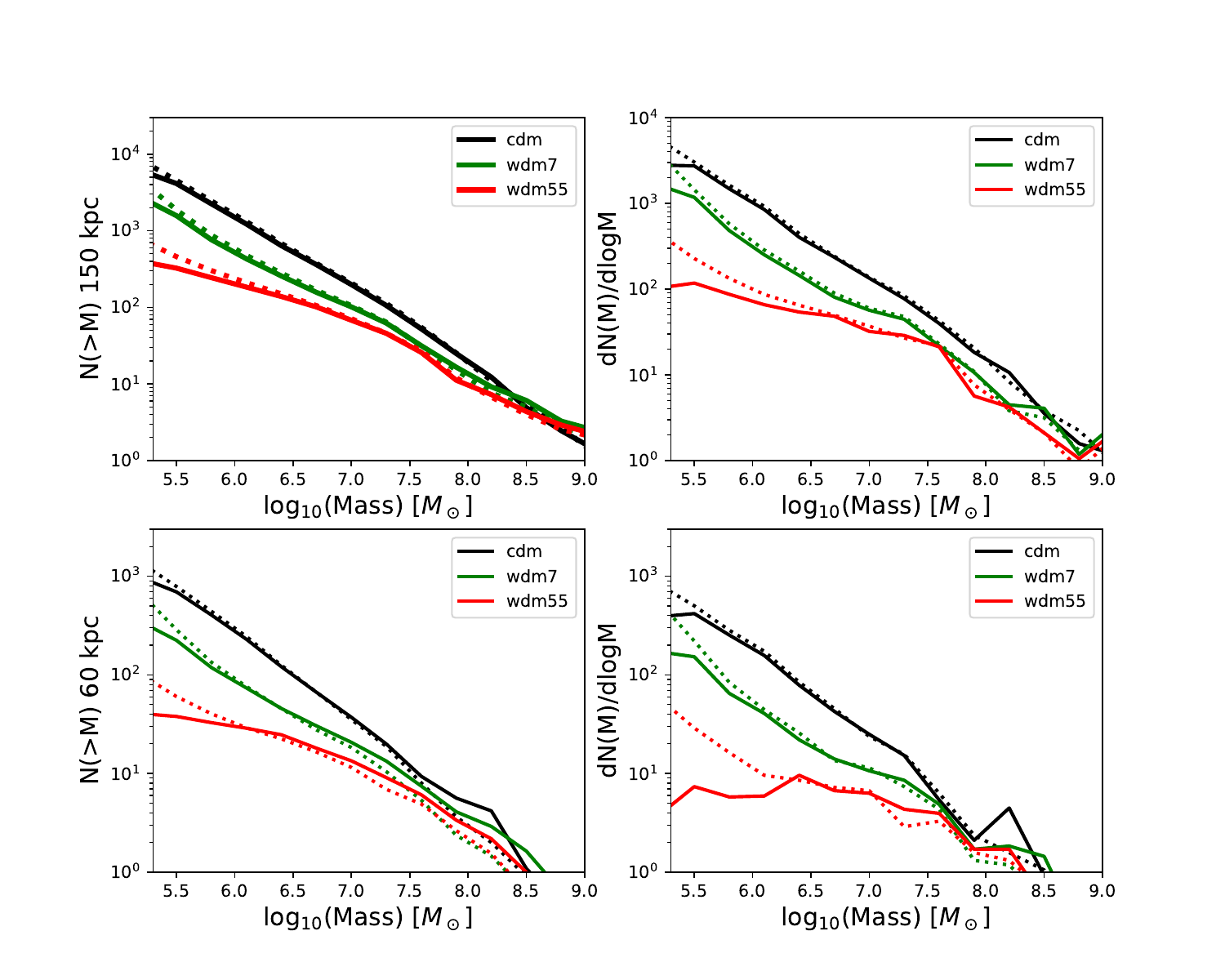}
\end{center}
\caption{
The subhalo mass (left panels) and differential mass functions (right panels) in the CDM,  WDM 7keV and WDM 5.5 keV simulations as measured with AHF (solid) and ROCKSTAR (dotted). The mass functions are measured every 0.2 Gyr over the last 2 Gyr and plotted as an average. The upper panels are measured within 150 kpc, 60 kpc in the lower panels. 
}
\label{fig_submf}
\end{figure}

\begin{figure}
\begin{center}
\includegraphics[angle=0,scale=0.40,trim=40 0 0 30, clip=true]{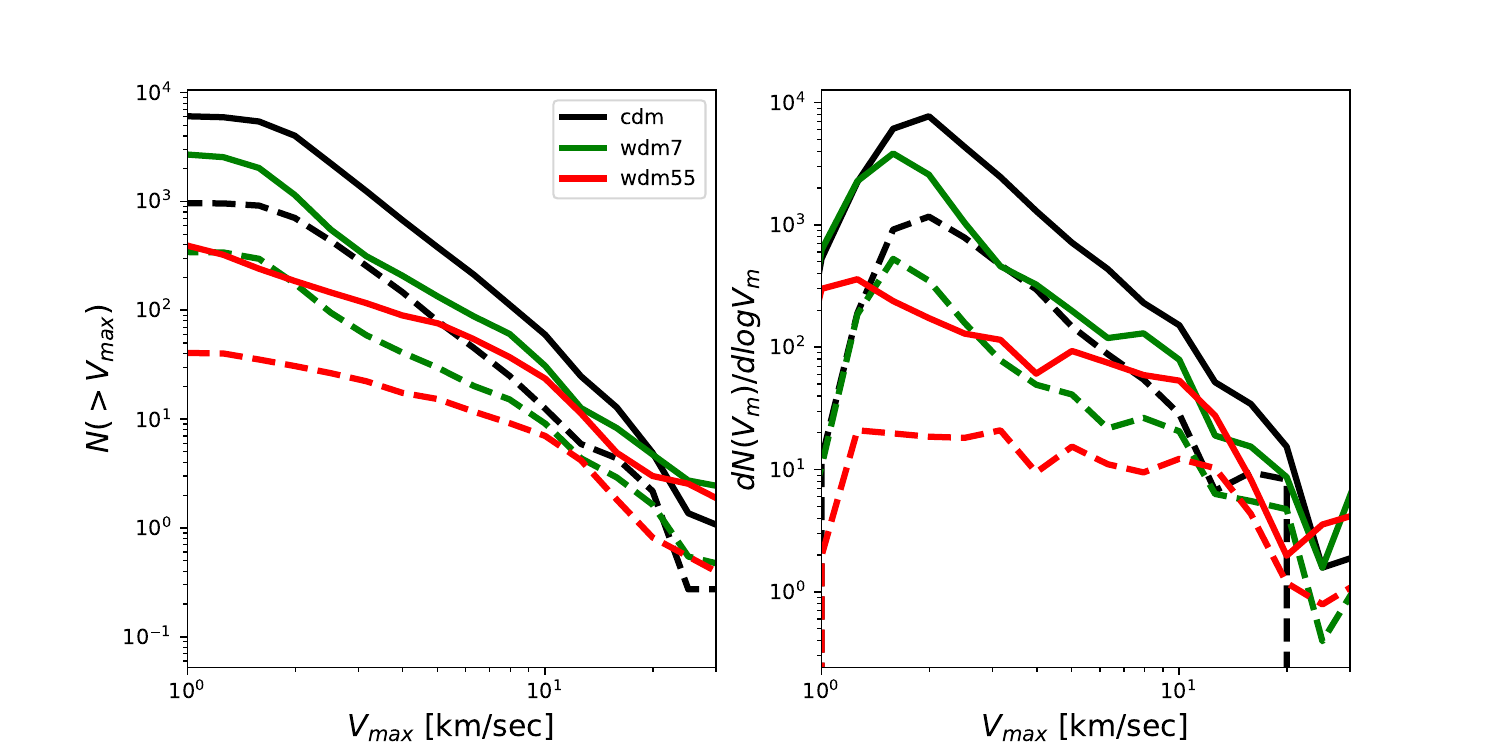}
\end{center}
\caption{
Subhalo numbers (cumulative left, differential right) with the maximum of the circular velocity of the halos, $V_{max}$, as measured with AHF.  The solid lines are the measurements within 150 kpc, the dashed lines within 60 kpc. 
}
\label{fig_subvf}
\end{figure}

Locating and characterizing subhalos within a larger halo depends on how the subhalos are defined \citep{SubHalos12,Symfind}. This paper uses the Amiga Halo Finder \citep{AHF1,AHF2}. An alternative is ROCKSTAR \citep{ROCKSTAR}. Figures~\ref{fig_submf} shows the mass functions and the numbers with velocity. The subhalo finders use the same minimum subhalo mass of $2\times 10^5 M_\odot$. ROCKSTAR as used here with the virial mass estimator on a single simulation epoch finds more low mass subhalos than AHF.  The two halo finders give similar results for masses greater than $10^{6.5} M_\odot$. The  maximum circular velocity of a subhalo, $v_{max}$, is more readily connected to dwarf galaxy kinematics and is shown in Figure~\ref{fig_subvf}. The subhalos are resolution limited below 2 \kms. The subhalos in the inner 60 kpc are the central regions of significantly more massive subhalos that were  tidally stripped as they orbited into the dense inner region of the main halo \citep{Errani22}.

\begin{figure}
\begin{center}
\includegraphics[angle=0,scale=0.54,trim=10 10 20 20, clip=true]{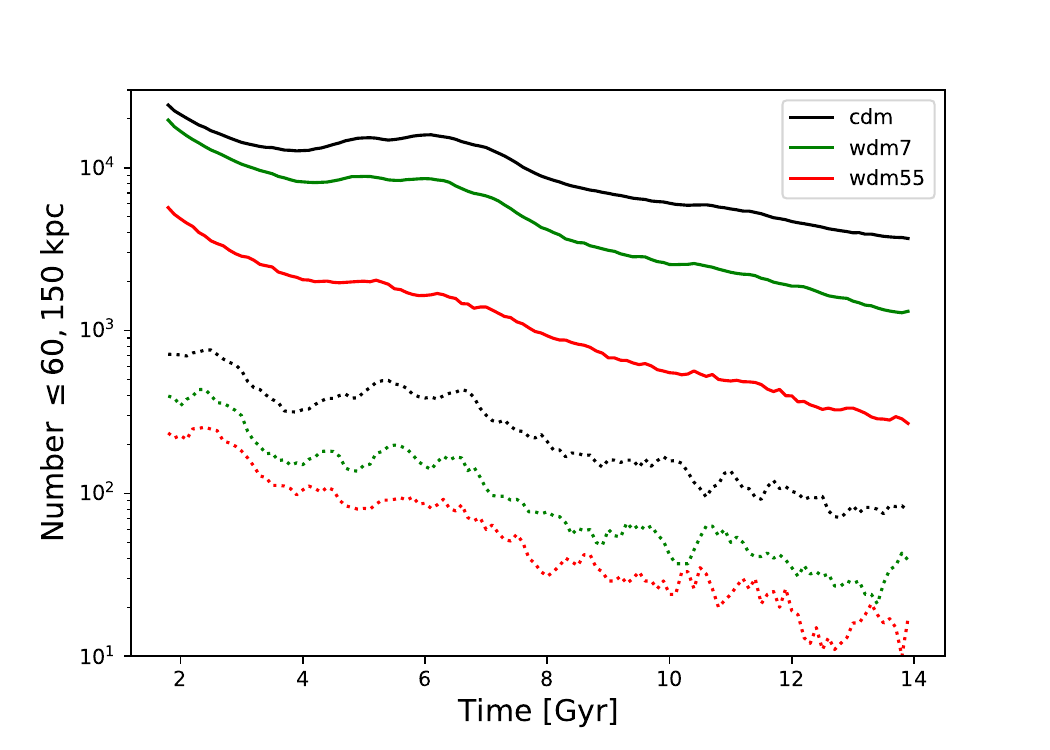}
\end{center}
\caption{
The numbers of subhalos with time. The solid lines are for all subhalos $\geq 10^{5.5} M_\odot$, $\gtrsim 2 \kms$ inside 150 kpc. The dotted lines are the numbers for the subhalos in the mass range of $10^{6.5}-10^{7.5} M_\odot$ inside 60 kpc, which dominate the stream velocity perturbations.
}
\label{fig_halont}
\end{figure}

The time evolution of the numbers of subhalos within fixed physical volumes is shown in Figure~\ref{fig_halont}. The solid lines show the numbers of AHF-halos more massive than $2\times 10^{5} M_\odot$ within 150 kpc. The lower dotted line shows the numbers of subhalos in the $10^{6.5-7.5} M_\odot$ range inside 60 kpc, where most of the thin streams orbit. The numbers of subhalos decline with time as tidal forces gradually shred them and dynamical friction moves them towards the center where the tides are stronger. Ongoing accretion and mergers bring in new subhalos which boost the numbers. A major merger occurs around 5 Gyr which temporarily increases the numbers of subhalos, as shown in Figure~\ref{fig_halont}. 

\section{Long Thin Streams\label{sec_streams}}

\begin{figure*}
\includegraphics[angle=0,scale=0.22,trim=130 130 130 130, clip=true]{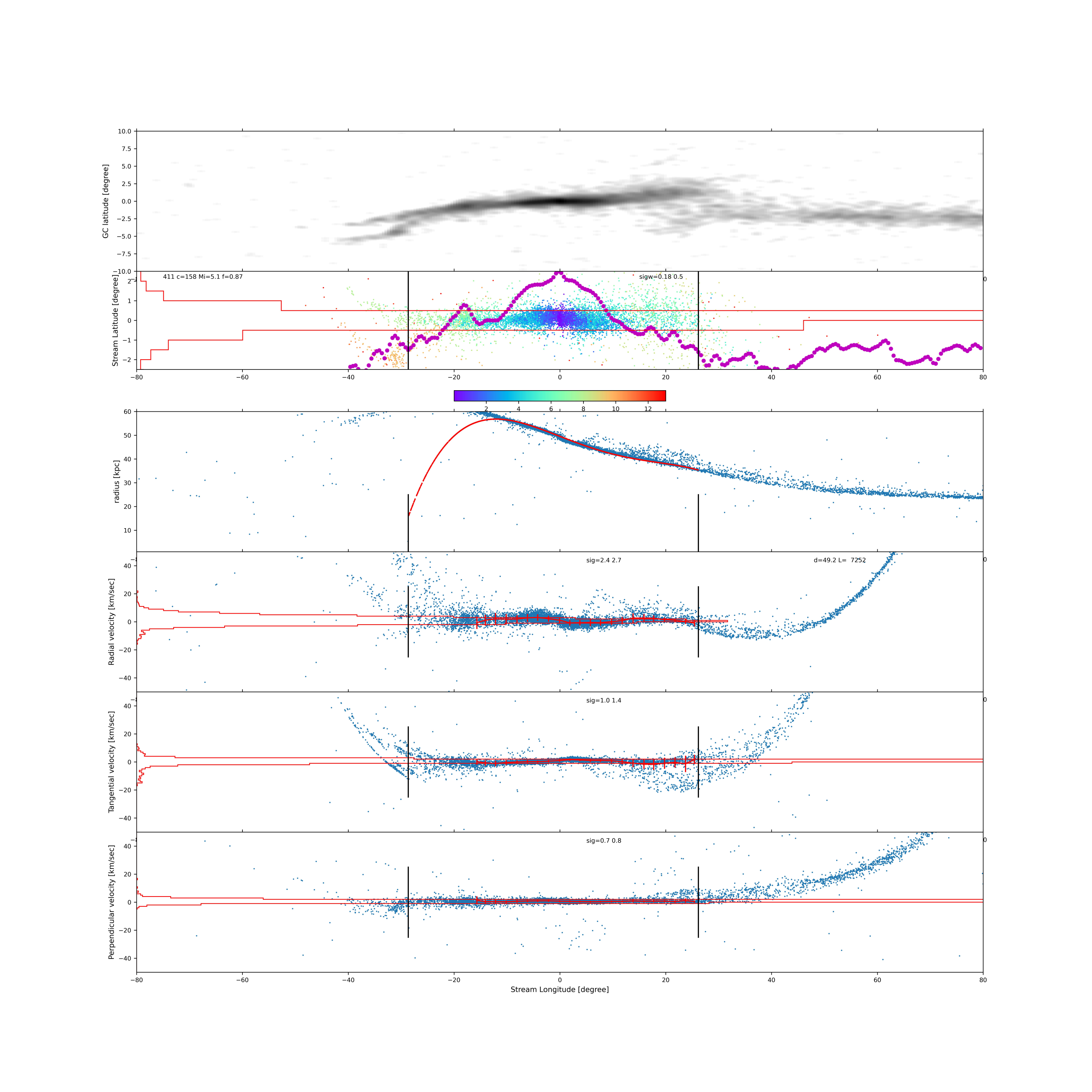}
\includegraphics[angle=0,scale=0.22,trim=130 130 130 130, clip=true]{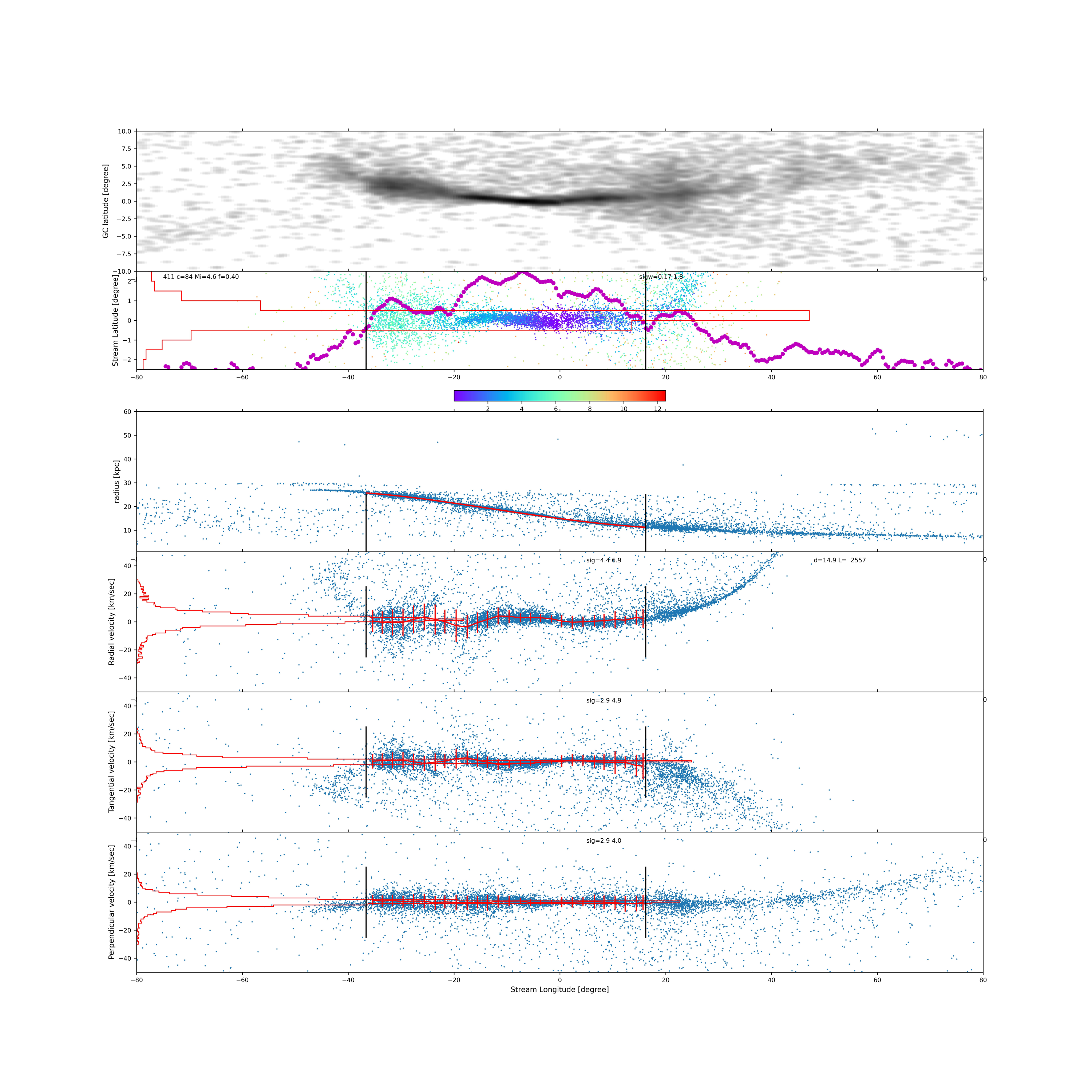}
\includegraphics[angle=0,scale=0.22,trim=130 130 130 130, clip=true]{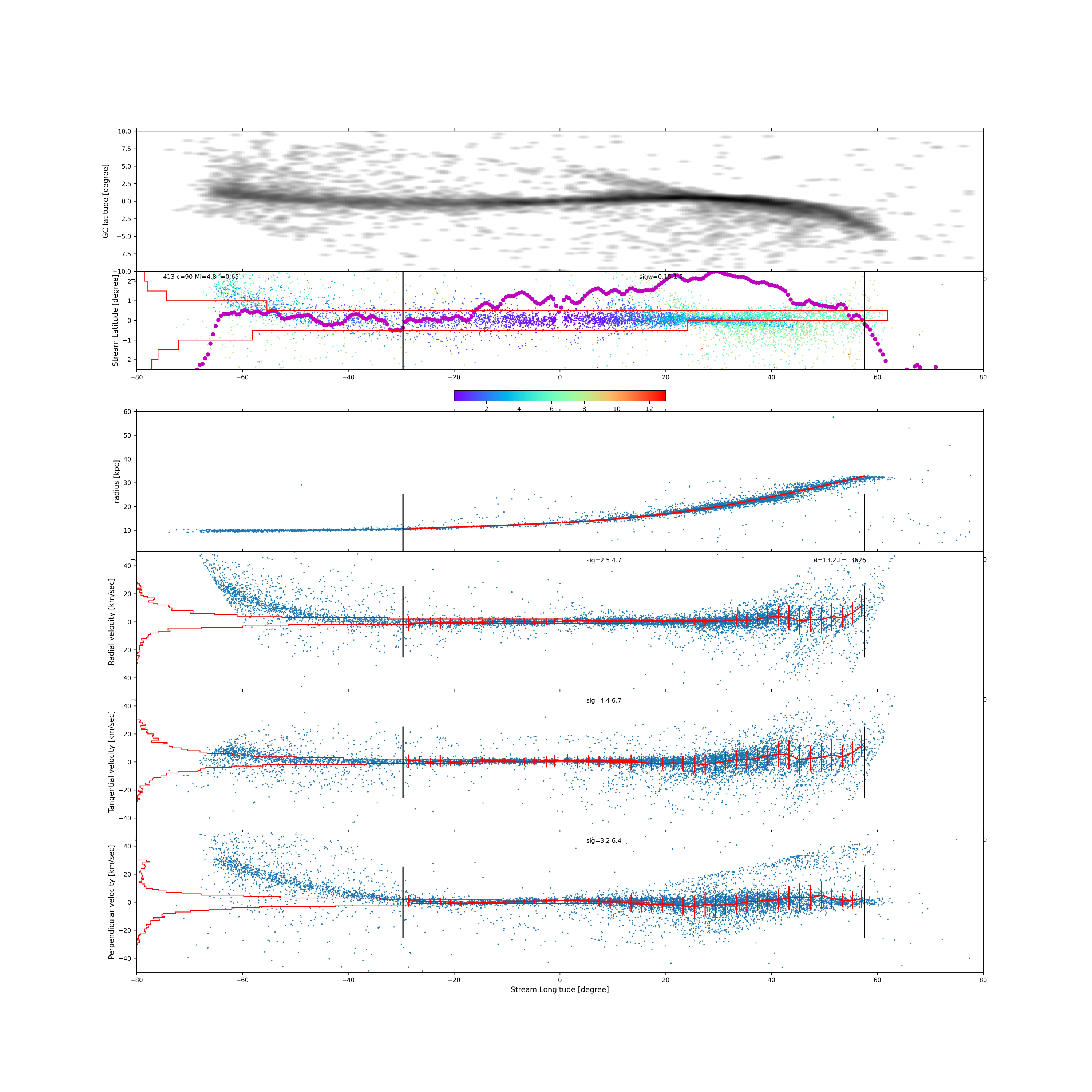}
\includegraphics[angle=0,scale=0.22,trim=130 130 130 130, clip=true]{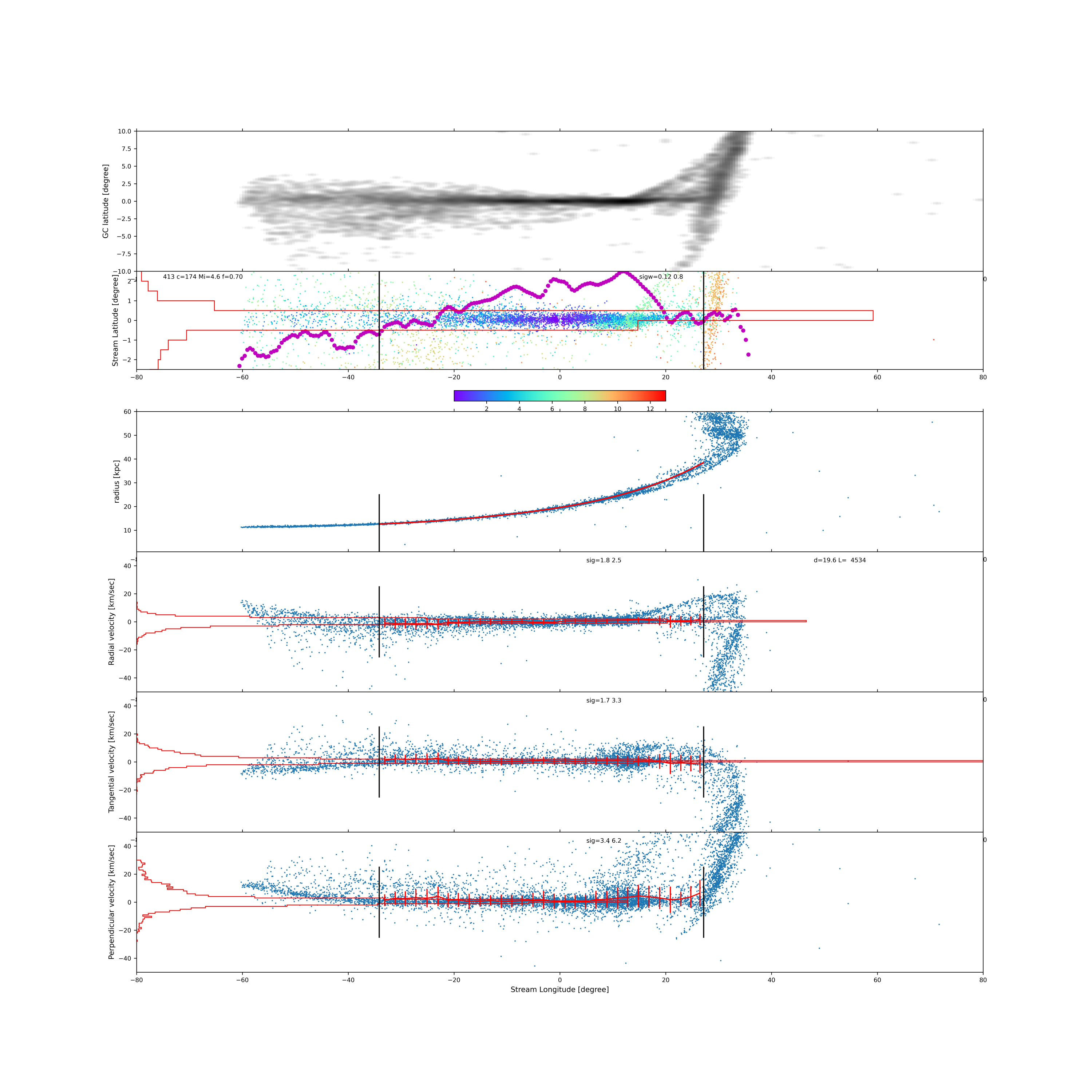}
\caption{
Thin, $\sigma_w \leq 0.2^\circ $, long, $\geq 40^\circ $, streams in CDM (top row) and WDM 5.5 kev  (bottom row).
Within each panel the sub-panels are:
(1, top) gray scale of the log surface density distribution with a 3 decade range for stream particles at all distances { in  the stream's galactocentric great circle coordinates}; 
(2) the individual stream particle positions after removal of a polynomial fit to  longitudinal variation of the streamline, with the surface density along the maximum as the magenta points normalized to the maximum with a range of 2 decades; (3) the particle radii with longitude,
(4-6) the radial, tangential and perpendicular stream particle velocities after subtraction of  polynomial fits to their azimuthal variation. The mean and standard deviation of the velocities are shown as the red error bars.
The horizontal histograms show the distributions about the zeros of the distribution. The black lines show the endpoints of the stream at a surface density at the longitudinal maximum of $20 M_\odot / \square^\circ$.
}
\label{fig_streams}
\end{figure*}

The globular clusters lose stars to tidal streams over their lifetime.  The stars released long ago are usually widely dispersed over the halo. Only the stars tidally pulled out into a stream in the last few Gigayear are in thin, high density segments.  The stream segments of greatest interest are the high surface density, thin, and long streams. Long streams  have the most interactions with subhalos along their length and thin streams are most sensitive to those interactions.  In the Milky Way the prototypes are  GD-1 \citep{GD1} and Pal~5 \citep{Pal5,Pal5_22deg}.  The simulations  result in  117 star clusters between 10 and 100 kpc, half of those within 60 kpc.  Every cluster has a stream.  The streams have a range of surface densities, widths and lengths depending on their orbital history in the initial high redshift dwarf halo which dissolves in the primary halo, and the subhalos encountered along the stream path.  

Measurement of stream properties follows observational procedures.  The streams are converted to galactocentric great circle coordinates defined with the progenitor location and velocity.  The progenitor is marked with a special $1 M_\odot$ particle initially located at the center. The center particle of each cluster is used to define a great circle coordinate system, with the particle at longitude zero and its angular momentum defining a reference plane for the stream.   The stream particles are then projected onto a cylindrical grid with pixels $0.5^\circ \times 0.1^\circ$, (width-height), then filtered with a 2D Gaussian of width 1.5 grid elements in both directions to give a density of the stream with latitude as a function of longitude, as shown in the top sub-panels of Figure~\ref{fig_streams}. The same procedure is used to put the particle radii and the radial, tangential and vertical velocities into radial and velocity grids with longitude. The positions of the particle density maxima in latitude, radius, and the 3 components of velocity are found. The stream endpoints are defined as where the centerline density drops below some minimum value (usually 20 $M_\odot$ per square degree, as measured on the filtered  $0.5^\circ \times 0.1^\circ$ grid). Or, if the stream centerline, as defined by the stream latitude of the highest projected density at every stream longitude, jumps by more than $2^\circ$ or if the radial or tangential velocity along the stream jumps more than 30 \kms, although these two discontinuity criteria do not usually prevail. The locations of the center lines are then fit with a fourth order polynomial in stream longitude. The particles that are close to these centerlines are then selected for measurement of the angular width and velocity spread. The first measure of closeness is the angular distance from the stream centerline \citep{CA23}. Here we use $\pm3^\circ$ to be within observational feasibility and fully capture the non-Gaussian wings of the velocity profile. We also require stars to be within 30 \kms\ of the stream center line in velocity and 5 kpc in radius. These criteria typically find about 5000 star particles close to a stream. All these parameters are reasonable, but illustrative, and can be adjusted to any practical observational program. 

The projected angular width of the core of a stream is a basic property, important for finding a stream on the sky and requires no kinematic data for its measurement. Most currently known  streams have an easily measured full width at half maximum (FWHM). For a Gaussian of width $\sigma_w$, the FWHM = $2.355\sigma_w$. For each simulation stream $\sigma_w$ is calculated as the $2\sigma$ clipped value using all star particles along the segment of the stream between the endpoints. Streams selected for detailed study are required to be longer than $40^\circ$, as seen from the center of the main halo. Imposing a maximum width of $0.2^\circ$ leaves 3 streams in the CDM simulation and 5 in  the 5.5 keV WDM. Relaxing the maximum allowed width to $0.3^\circ$ gives 10 and 12 streams in the same two simulations. Two representative CDM and two WDM 5.5 keV streams are shown in Figure~\ref{fig_streams}.

\begin{figure}
\begin{center}
\includegraphics[angle=0,scale=0.22,trim=70 93 85 60, clip=true]{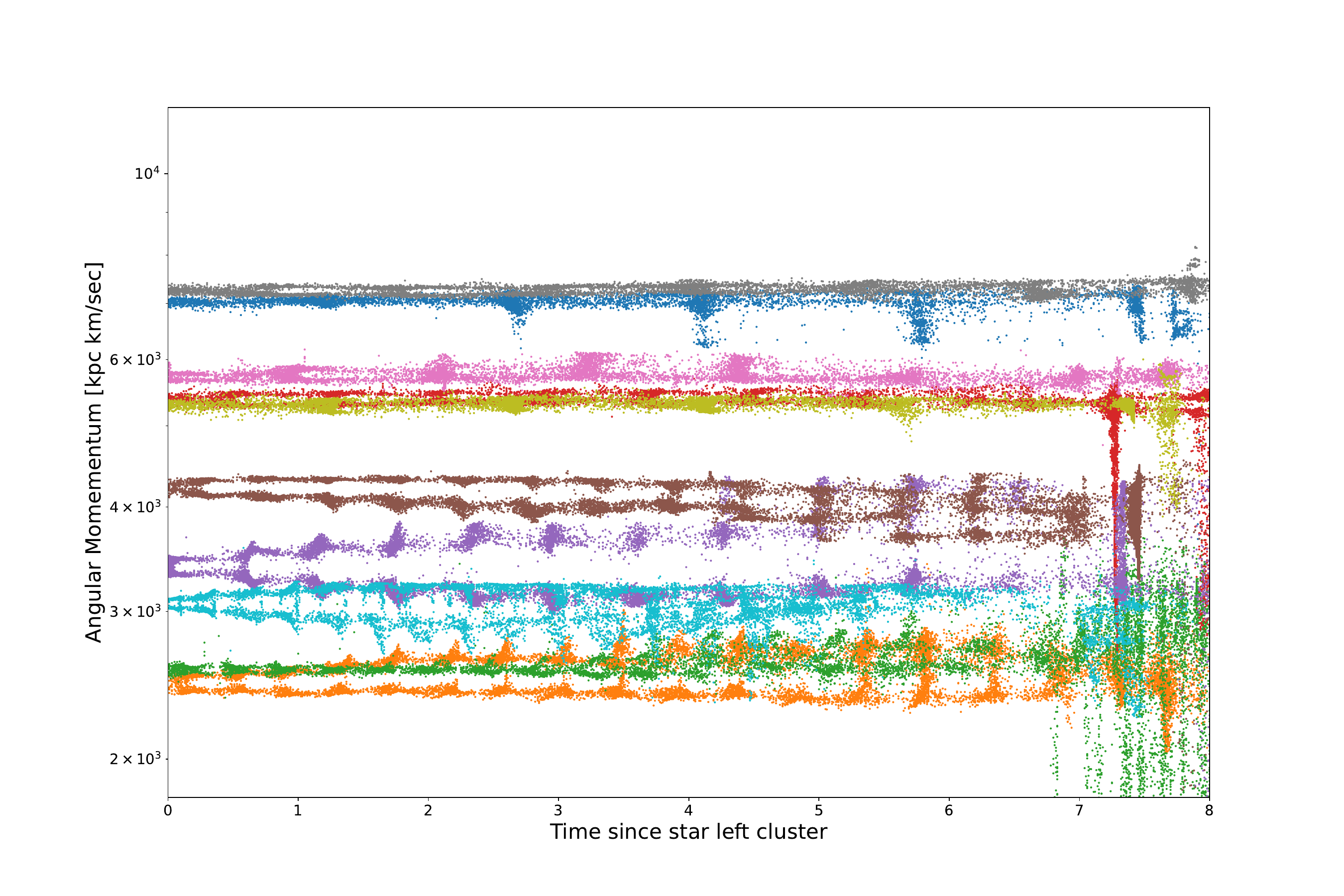}
\includegraphics[angle=0,scale=0.22,trim=70 60 85 80, clip=true]{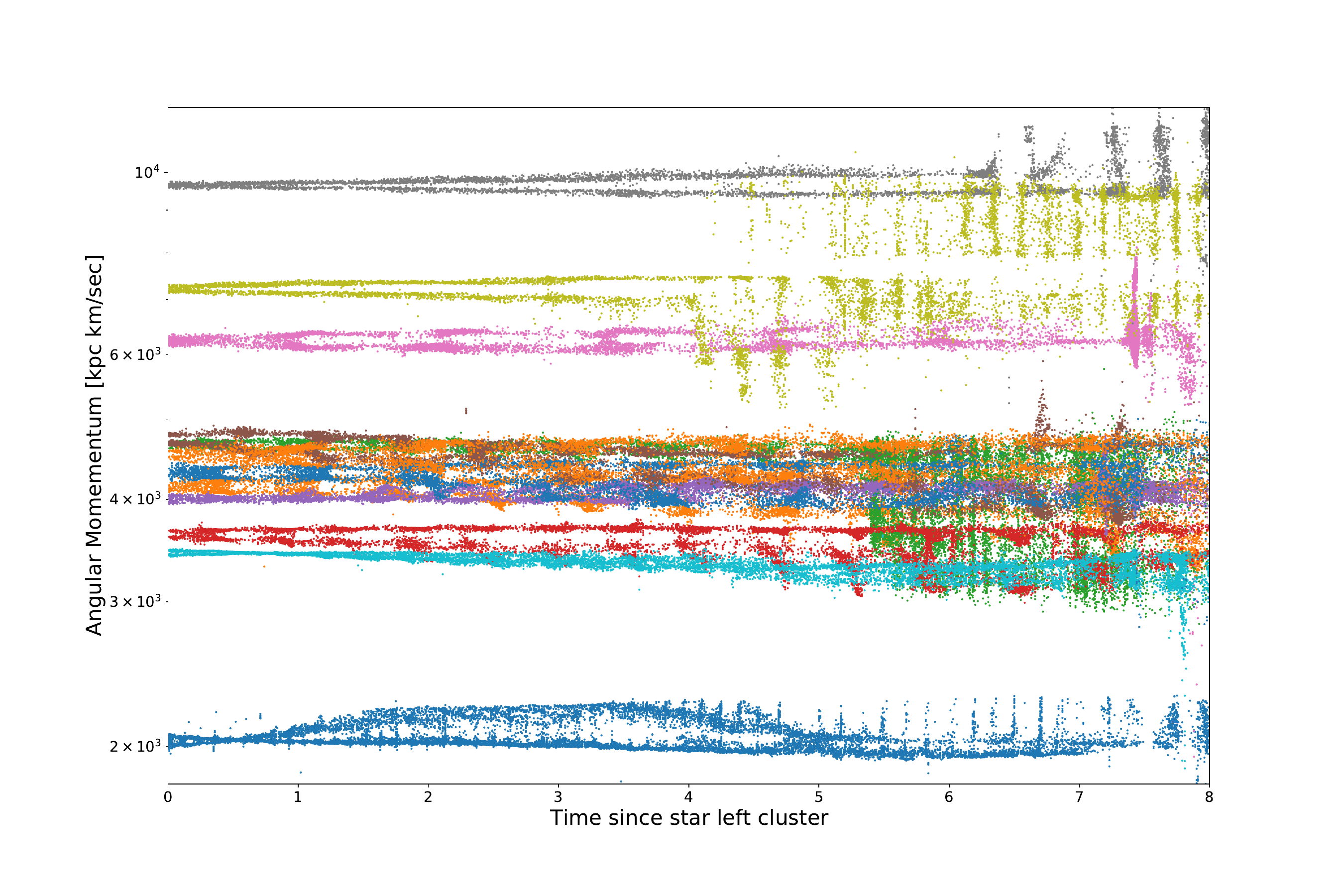}
\end{center}
\caption{
The angular momenta of stream star particles measured at the end of the simulation versus the length of time that the particles have been in the stream. The streams are the thin, long streams in the CDM simulation (top) and the 5.5 keV WDM (bottom).
}
\label{fig_NCLtime}
\end{figure}

The stream density on the sky is shown in the top sub-panel of each stream plot of Figure~\ref{fig_streams}. The sky density gray scale is logarithmic over three decades with the peak density scaled to black. The second sub-panel shows the individual star particles in each stream after removing the variations around the stream equator with a 4th order polynomial. Note that the range in stream latitude is 4 times less than in the top plot. The vertical black lines mark the location of the endpoints as defined above. The horizontal histogram shows the distribution of particles about the centerline. The magenta points give the logarithm of the stream density along the centerline, with a range of 2 decades. The third sub-panel has the galactocentric radii of the particles in kpc.  The red line shows the fitted polynomial. The fourth sub-panel from the top shows the residual galactocentric radial velocities of the stream particles after removing the fitted trend with longitude.  The horizontal histogram is the distribution of the velocities within the selected region. The mean and standard deviation of the velocities are also displayed as error bars in $2^\circ$ bins along the stream. The two velocity components in the plane of the sky, tangential and perpendicular to the polynomial corrected mean stream are shown in the bottom two sub-panels, respectively. 

Star streams are a collection of stars on nearby orbits \citep{Binney08}. Figure~\ref{fig_NCLtime} shows the instantaneous angular momentum of star particles measured at the end of the simulation as a function of the time since they left the cluster for the $\sigma_w\leq 0.3^\circ$ streams in the CDM and WDM 5.5 keV simulations. Angular momentum is not a conserved quantity in these aspherical halos but varies relatively little around an orbit for most stream stars. The spread of angular momenta in a stream means that stars will have different orbital frequencies and will shear with respect to one another along a stream. 

Movies of the simulations and streams are available at \href{https://www.astro.utoronto.ca/~carlberg/streams/411/movies}{CDM}, and \href{https://www.astro.utoronto.ca/~carlberg/streams/413/movies}{WDM 5.5 keV}. Halo globular clusters are old star systems that have lost stars to the evolving galactic halo over their lifetime.  Even at late times when the potential is relatively slowly and smoothly varying  the high density, thin sections of the streams shrink, stretch, and change morphology around their orbits. Encounters with subhalos usually do not have easily identifiable signatures because of the complex phase space structure of streams. 

\section{Semi-Analytic Stream Dynamics\label{sec_dynamics}}

\begin{figure}
\begin{center}
\includegraphics[angle=0,scale=0.22,trim= 0  30 42 40, clip=true]{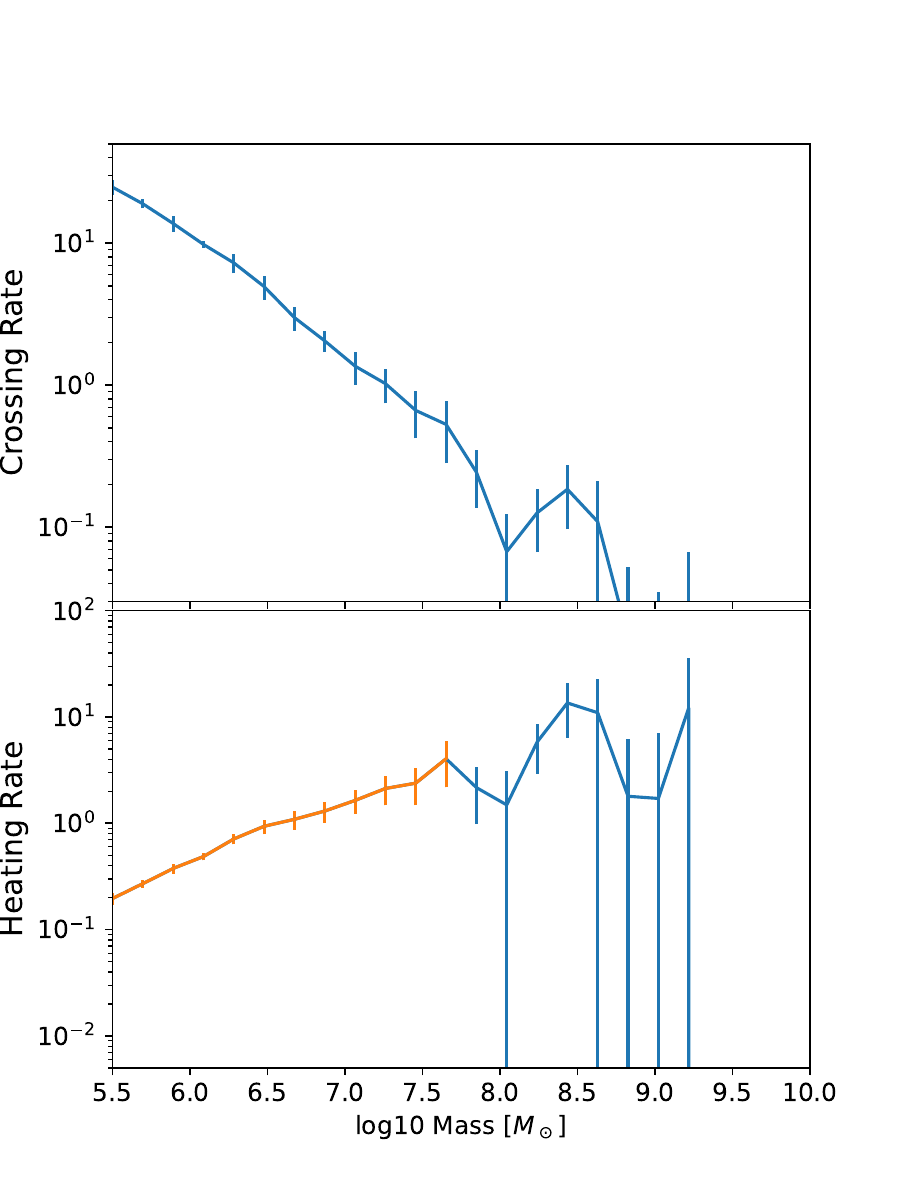}
\includegraphics[angle=0,scale=0.22,trim=50 30 42 40, clip=true]{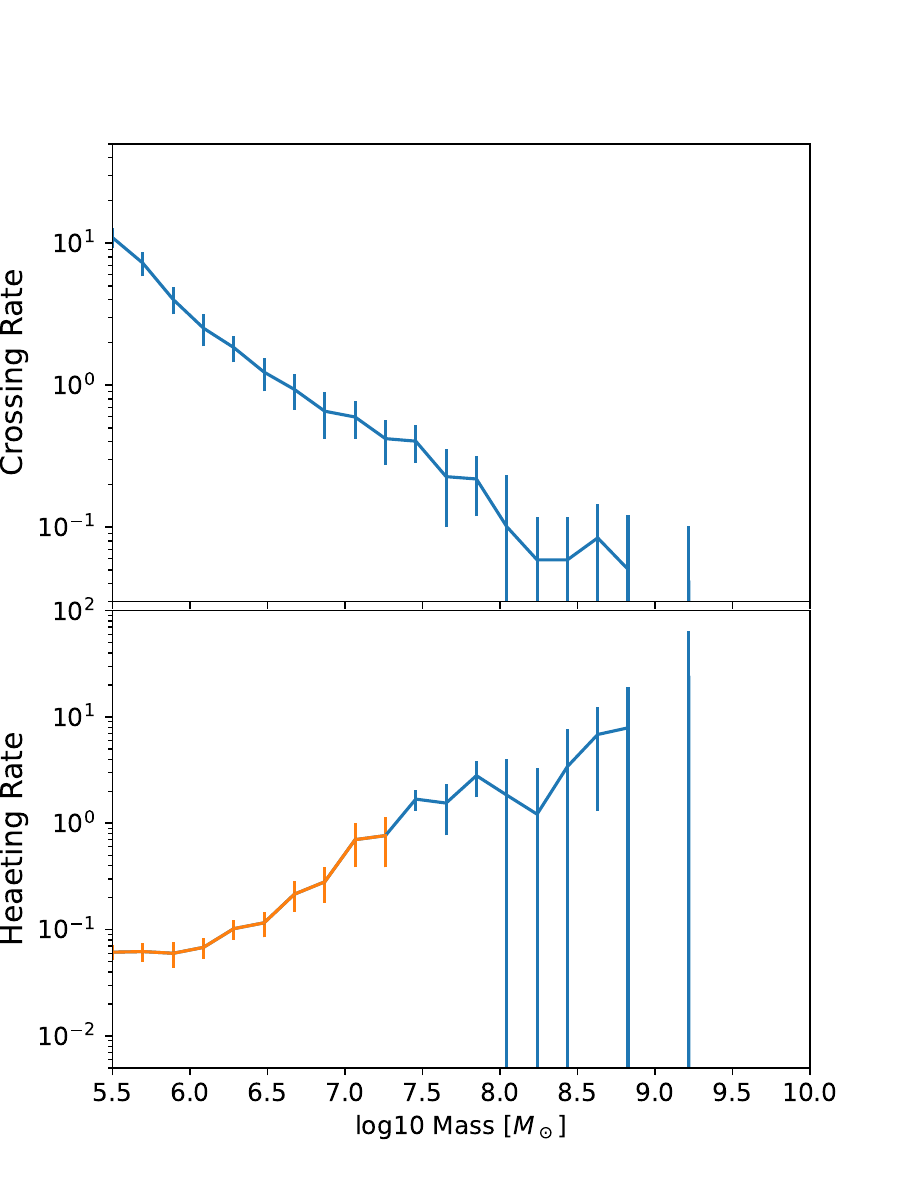}
\includegraphics[angle=0,scale=0.22,trim=50 30 42 40, clip=true]{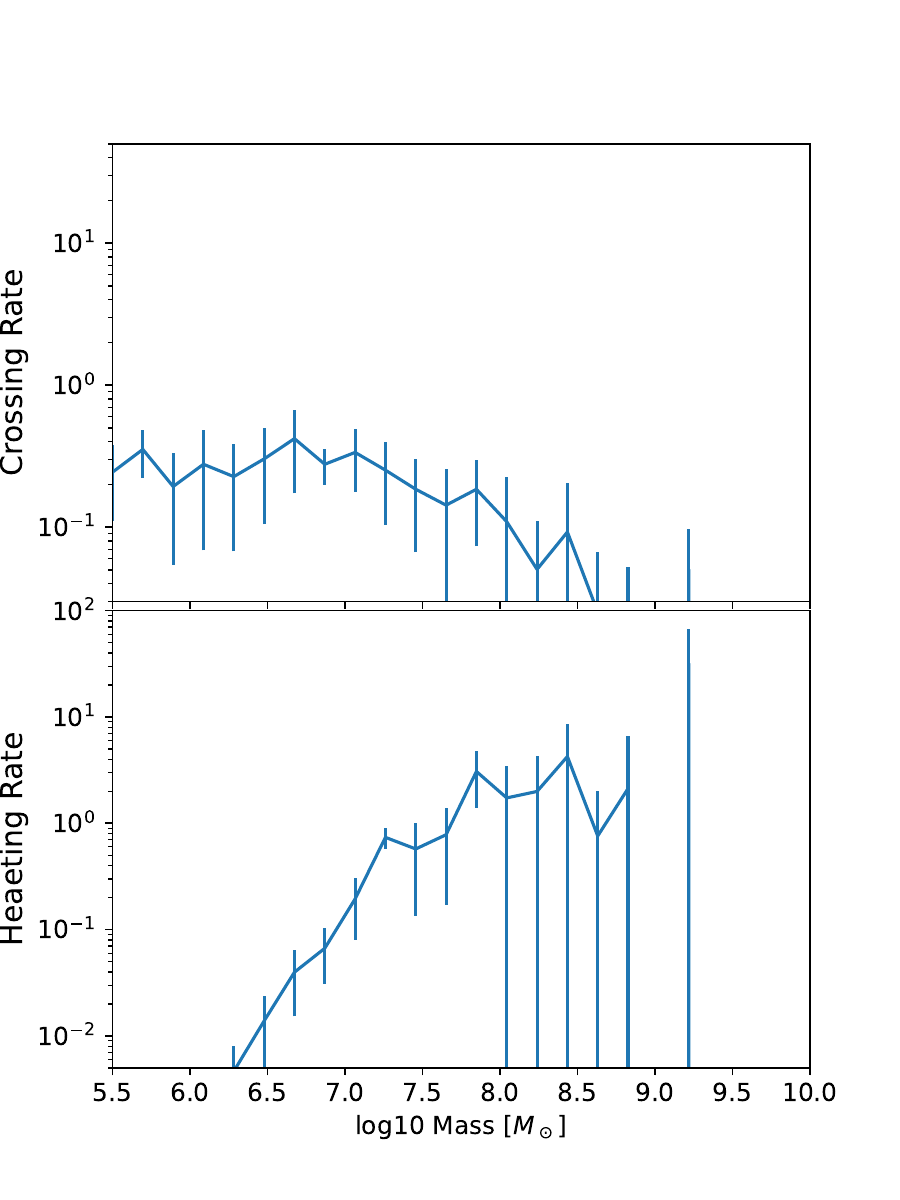}
\end{center}
\caption{
Semi-analytic estimates of the late time rate at which subhalos  cross a  10 kpc length of stream in 4 Gyr in CDM (left) and WDM (7 and 5.5 keV, middle and right, respectively). The bottom panel shows the velocity heating rate, $d\delta v^2/dt$. The orange portion of the lines in the lower panels is where there is one or more subhalo stream crossing.
At 7 Gyr, approximately redshift 1, the rates are about 4 times higher.
}
\label{fig_heatnum}
\end{figure}

The approximate calculation of the dynamics of sub-halo stream interactions in \citet{CA23} is updated with the subhalo measurements here to provide context for the velocity modeling. The rate per unit length at which sub-halos cross a 10 kpc length of stream in 4 Gyr in the inner 60 kpc of  the halo is shown in the top panels of Figure~\ref{fig_heatnum}. There is only one $\gtrsim 10^7 M_\odot$ subhalo crossing of an average 10 kpc segment of a stream in 4~Gyr. That is,  the more massive subhalos affect regions of a stream that usually do not overlap. The lower mass subhalo encounters are sufficiently frequent to produce a quasi-random walk in velocity which increases the velocity spread. In the 5.5 keV WDM model the subhalo numbers are so low that subhalo encounter rates are low for all masses. 

The rate of velocity spread is approximately,
\begin{equation}
{{d\delta v^2}\over{dt}} =4\pi^2 \sum_{M_s( \leq r)} N(M_s) {{ v_{max}(M_s)^2}\over{v_o V(r)}},
\label{eq_semiheat}
\end{equation}
which is updated from  \citet{CA23} to use the maximum of the circular velocity of a subhalo, $v_{max}\simeq \onehalf\sqrt{GM_s/a_s}$, rather than a subhalo mass and scale radius. The sum is over all the subhalos within the volume of interest, which we usually take to be 60 kpc. In principle the heating should be integrated over the relative velocity distribution of the streams and the subhalos. Here the relative orbital velocity $v_o$ is set  to $\sqrt{2} \sigma_{3D}$ where the 3D velocity dispersion for the main halo is measured to be 210 \kms, so $v_o\simeq 300 \kms$ is used in Equation~\ref{eq_semiheat} to calculate the heating rate per unit length in the simulations shown in Figure~\ref{fig_heating} at redshift zero. At earlier times the heating will increase in proportion to the subhalo numbers as shown in Figure~\ref{fig_halont}. In the CDM simulation the estimated heating over 4 Gyr per 10 kpc of length for subhalos that encounter streams more than once in the 4 Gyr is about 1.2 \kms. The velocity changes in response to a 300 \kms\ encounter are small, typically $1 \kms$ for a $10^{7.5} M_\odot$ subhalo.  The spread of orbits in the highly aspherical halo potential significantly augments the effects of subhalo encounters, as illustrated in \citet{Ngan16} with orbital measurements in \citet{CA23}. 

\section{The Stream Velocity Distribution\label{sec_fitting}}

\begin{figure}
\begin{center}
\includegraphics[angle=0,scale=0.52,trim=10 0 0 40, clip=true]{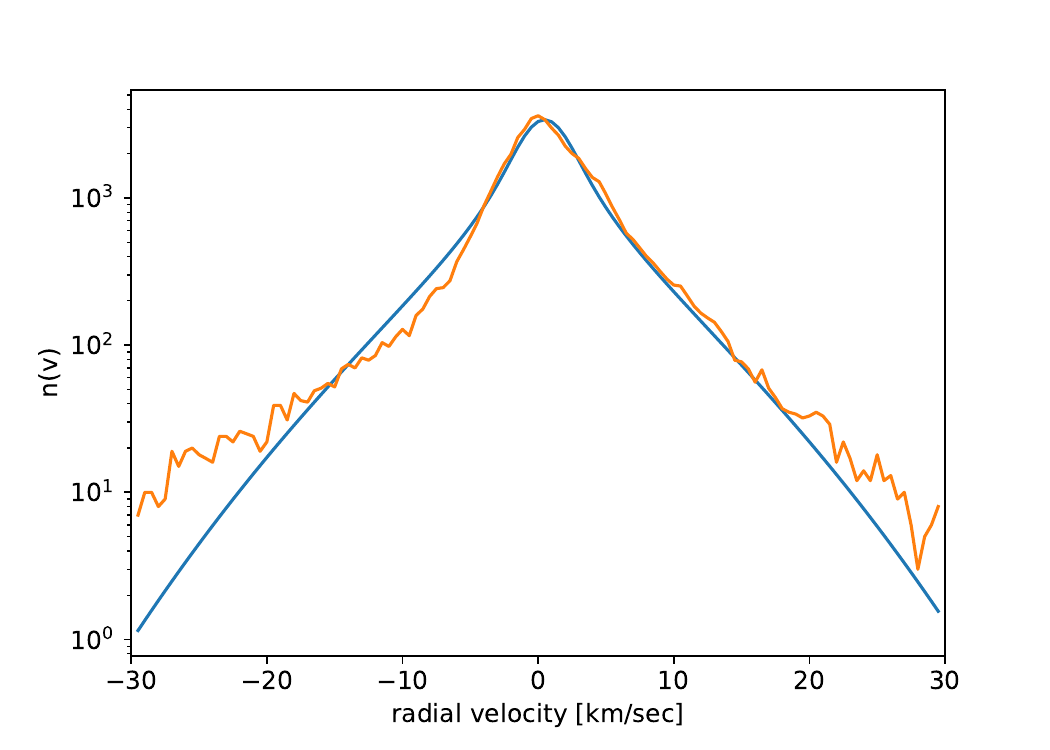}
\end{center}
\caption{
The CDM simulation radial velocity distribution fitted with the dynamical model, Equation~\ref{eq_fvdyn}.
}
\label{fig_fvdyn}
\end{figure}

\begin{figure}
\begin{center}
\includegraphics[angle=0,scale=0.28,trim=0 0 0 20, clip=true]{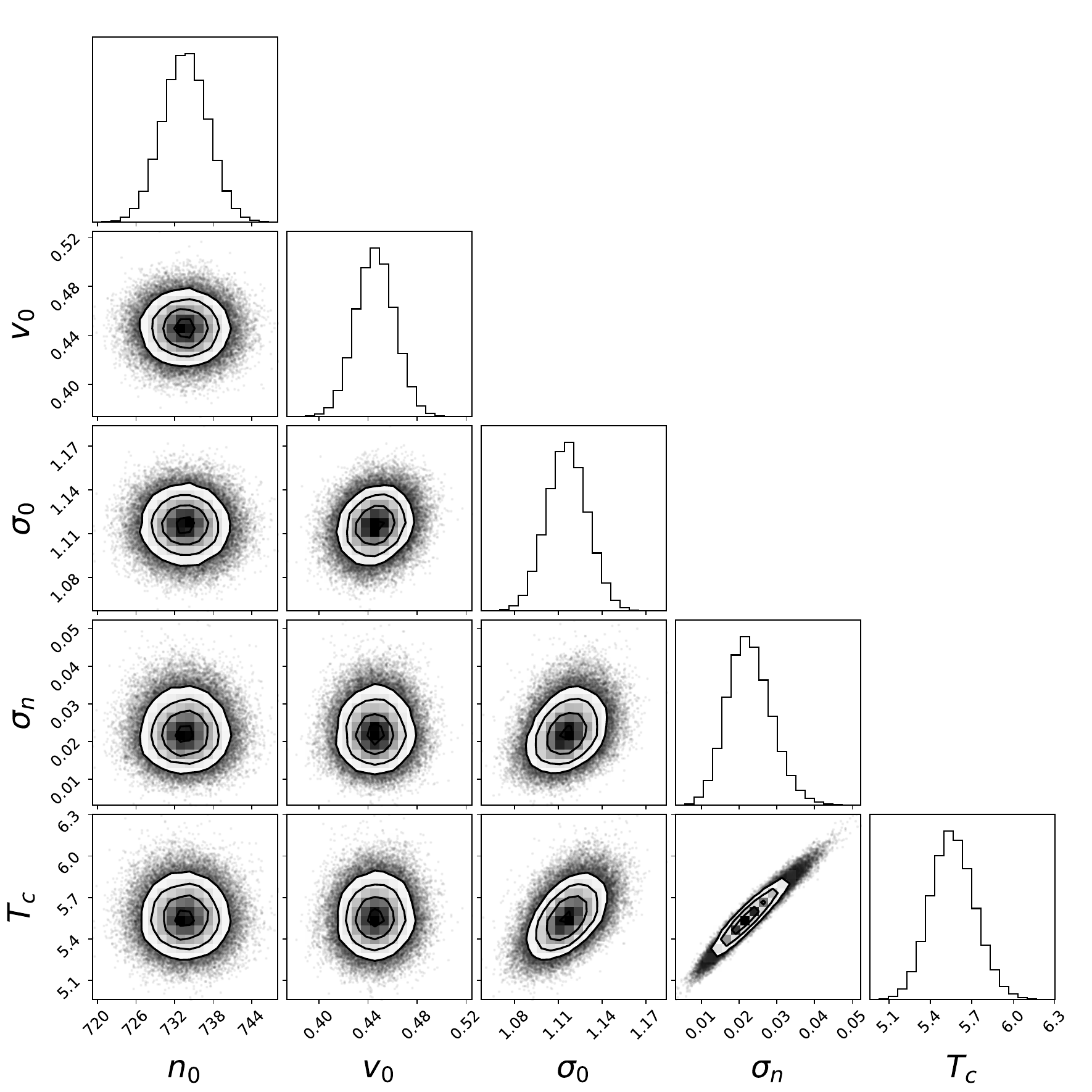}
\end{center}
\caption{
Corner plot for the dynamical fit,  Equation~\ref{eq_fvdyn}, of the stream radial velocity distribution in the CDM simulation shown in Figure~\ref{fig_fvdyn}.
}
\label{fig_corner_dyn}
\end{figure}

The distribution of stream star radial velocities around the local mean has the most straightforward dependence on subhalo numbers. An observational benefit is that line of sight velocities, often the dominant component of the radial velocity, can be measured spectroscopically to large distances, eventually to other galaxies of the local group, whereas the tangential components require astrometric measurements which are currently available within the inner halo only.  The selection of thin streams leads to streams with similar density and velocity distributions in the stream latitude direction. The tangential random velocity distribution also shows little dependence on subhalo numbers.  The selection procedure used here tends to find stream segments near pericenter where the mean radial velocities are relatively low, and the mean tangential velocities are high.   The velocities here are in the galactocentric frame.

The stars are pulled away from their parent cluster with a velocity dispersion characteristic of the cluster envelope, $\sigma_0 \simeq 1$ \kms\ after which random encounters with subhalos incrementally increase the velocity dispersion with time.  Averaging the velocity distribution over the visible length of the stream discards any information about the length of time that a star has been in the stream. In addition to the heating from subhalos the orbits of the stars disperse in the highly aspherical potential in a process that appears chaotic \citep{Ngan15,Ngan16,CA23}, hence, modeled as an exponential in time \citep{BT08}. Consequently a dynamically motivated model for the increase in velocity dispersion with time, { $\tau=t_0-t$, which is zero at the current epoch, $t_0$.
\begin{equation}
\sigma_v^2(\tau) = [\sigma_0^2 +\sigma_n (\tau/1~{\rm Gyr })^{ n_w}]e^{\tau/T_c}.
\label{eq_sigdyn}
\end{equation}
If the numbers of subhalos were constant in time,  ${{d\delta v^2}/{dt}}$ is a constant, see Eq.~\ref{eq_semiheat}, and $\sigma_v^2 \propto \tau$. However the density of subhalos rises into the past, as shown in Figure~\ref{fig_halont}.  Because the rate of increase in halo numbers is greater than zero but below a linear rise, we expect $n_w$ to be in the range of 1-2.} The resulting model velocity distribution function is the integral over time, here implemented as a sum over intervals of 1 Gyr to $T_{max}=11$~Gyr,
\begin{equation}
f(\delta v_r) = n_0 \sum_{\tau=0}^{T_{max}} \exp{\left[-\onehalf\frac{\delta v_r^2}{\sigma_v^2(\tau)}\right]} ,
\label{eq_fvdyn}
\end{equation}
where $\sigma_v(\tau)$ is from Equation~\ref{eq_sigdyn}. The lower limit of $\tau=0$ assumes that the star cluster does not completely dissolve, which is generally the case for the clusters in these simulations. A drawback of this model is that the resulting $\sigma_n$ and $T_c$ scale with the chosen $T_{max}$. However, the main purpose of this model is to show that velocity width evolution based on subhalo scattering provides a reasonable description of the wings of the velocity distribution.

\begin{figure}
\begin{center}
\includegraphics[angle=0,scale=0.7,trim=0 40 0 60, clip=true]{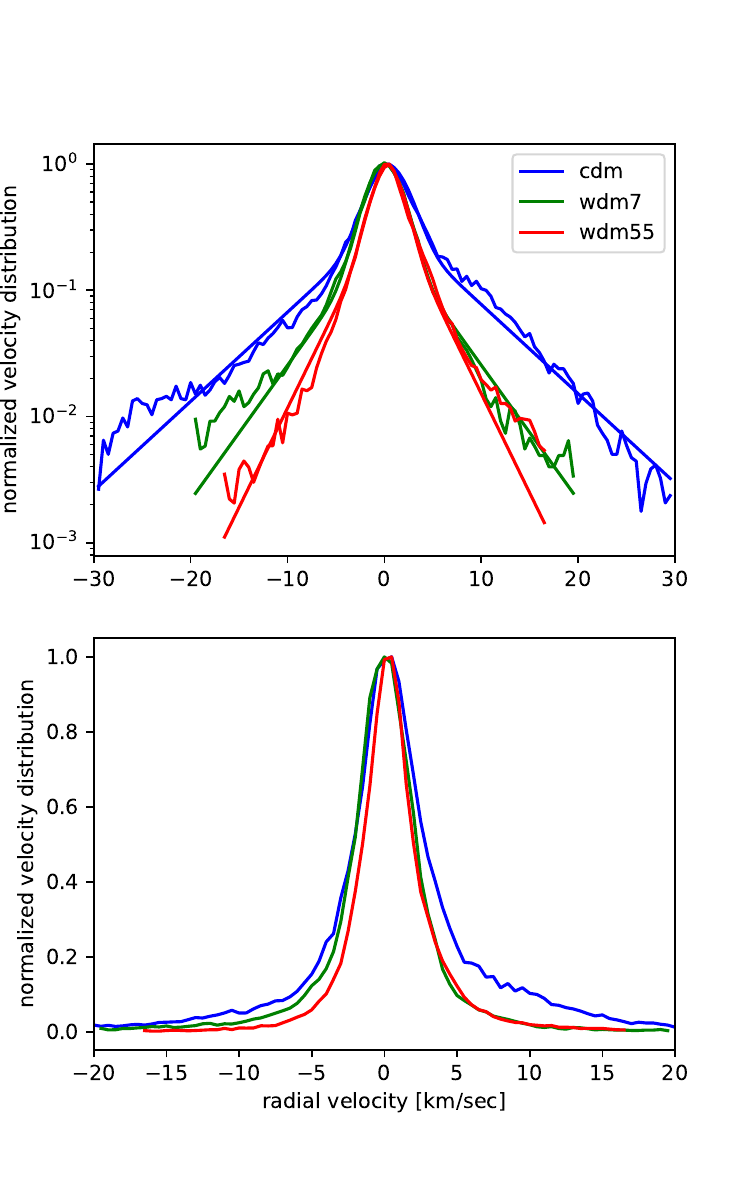}
\end{center}
\caption{
The CDM and WDM radial velocity distribution (logarithmic, top and linear, bottom) and their model fits (top panel only).
The velocity distribution has a yet wider extended skirt beyond the fitted ranges. 
}
\label{fig_fvge}
\end{figure}

\begin{figure}
\begin{center}
\includegraphics[angle=0,scale=0.28,trim=0 5 0 20, clip=true]{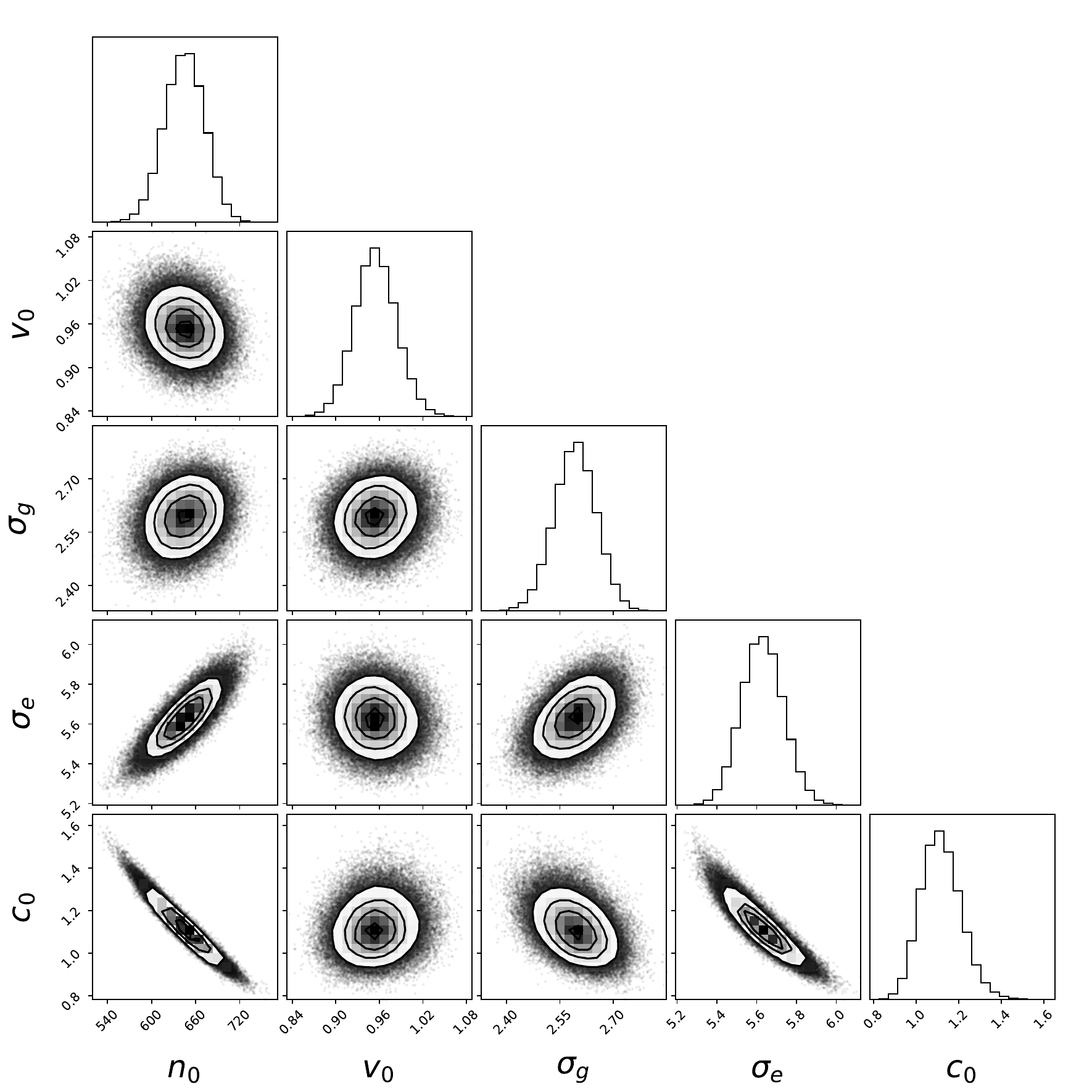}
\end{center}
\caption{
Corner plot for the Gaussian plus exponential fit,  Equation~\ref{eq_fvge}, of the stream radial velocity distribution in the CDM simulation shown in Figure~\ref{fig_fvge}.
}
\label{fig_corner_ge}
\end{figure}

The  radial velocity distribution for the selected CDM streams and its MCMC model fit \citep{emcee} to Equation~\ref{eq_fvdyn} is shown in Figure~\ref{fig_fvdyn}. The fit is quite good over a factor of about 30  in dynamic range but is not sufficiently wide beyond about 15 \kms. The corner plot for the fitting parameters is shown in Figure~\ref{fig_corner_dyn}, revealing that the dynamical heating $\sigma_t$ and the orbital diffusion $T_c$ are correlated (along with the externally fixed $T_{max}$), although both are reasonably well constrained in their marginalized distributions. When applied to the WDM velocity distributions the dynamical model of Equation~\ref{eq_fvdyn} produces comparable results for the 7 keV model but underestimates the extended wings of the 5.5 keV model. Nevertheless, the ability of the dynamical model to provide a fit with reasonable parameters helps validate the underlying idea that the velocity distribution within the inner $\approx$20 \kms\ of the stream is largely the result of subhalo impacts and orbit diffusion in the aspherical potential.

A simple and robust empirical velocity distribution fitting function is more useful than the dynamical model to compare simulation results to each other and to observational data.  The logarithm of the velocity distribution in Figure~\ref{fig_fvdyn} suggests a model with a Gaussian core and exponential wings, with widths of $\sigma_g$ and $\sigma_e$, respectively,
\begin{equation}
f_{ge}(v) = n_0\left[ \exp{ \left( -\onehalf \frac{v^2}{\sigma_g^2} \right) } 
	+ c_0 \exp{\left(-\frac{\left|v\right|}{\sigma_e}\right)} 	\right].
\label{eq_fvge}
\end{equation}
The results of the MCMC fits are shown for the CDM and WDM long, thin, streams in Figure~\ref{fig_fvge}. Figure~\ref{fig_corner_ge} shows the corner plot for the CDM fit. This simple model function captures most of the range for the long, thin, streams in the CDM and WDM simulations.

\begin{figure}
\begin{center}
\includegraphics[angle=0,scale=0.72,trim=5 55 20 70, clip=true]{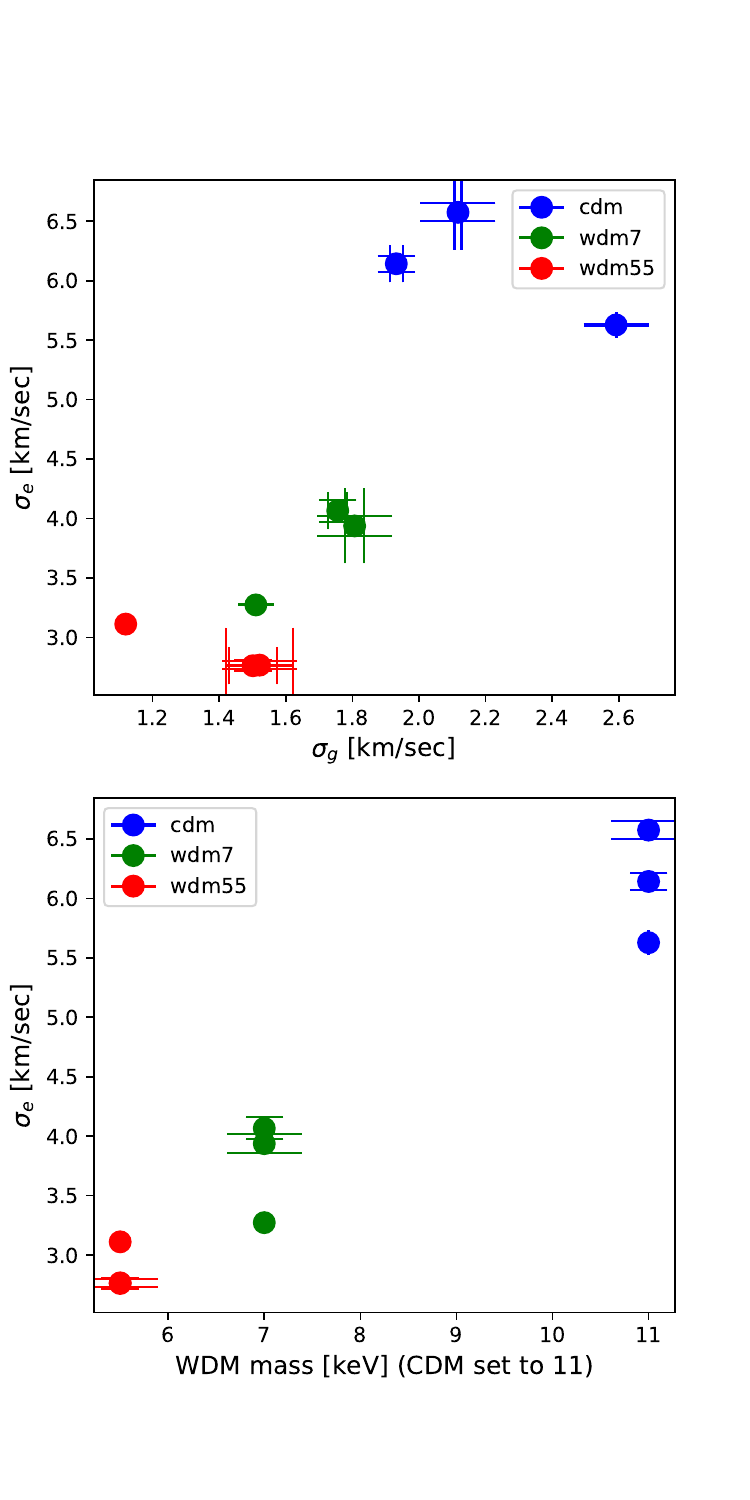}
\end{center}
\caption{
The MCMC derived $\sigma_g$ and $\sigma_e$ of the Gaussian plus exponential model for the radial velocities in CDM and WDM streams having widths $\sigma_w\leq 0.2^\circ$, $0.25^\circ$, and $0.3^\circ$ (indicated as caps of increasing lengths on the error bars),  longer than $40^\circ$. There are 3-12 streams in the averages.
}
\label{fig_paramsge}
\end{figure}

Streams selected with $\sigma_w$ of $0.2^\circ- 0.3^\circ$ are fit with the Gaussian plus exponential model with results shown in Figure~\ref{fig_paramsge}. The lower panel of Figure~\ref{fig_paramsge} shows that the exponential width of the radial velocity distributions is a good measure of the numbers of subhalos in the different dark matter models.  

The measurements here are from well resolved n-body simulations with low noise levels and no measurement uncertainties. The widths of the exponential component of the velocity distributions of  the WDM simulations, are 3 and 4 \kms, for 5.5 and 7 keV, respectively, whereas CDM has 6 \kms. The $\sigma_e$ are different enough that the models can be compared to practical observations with realistic velocity errors. 

\section{Observational Considerations\label{sec_errors}}

\begin{figure}
\begin{center}
\includegraphics[angle=0,scale=0.4,trim=50 0 20 0, clip=true]{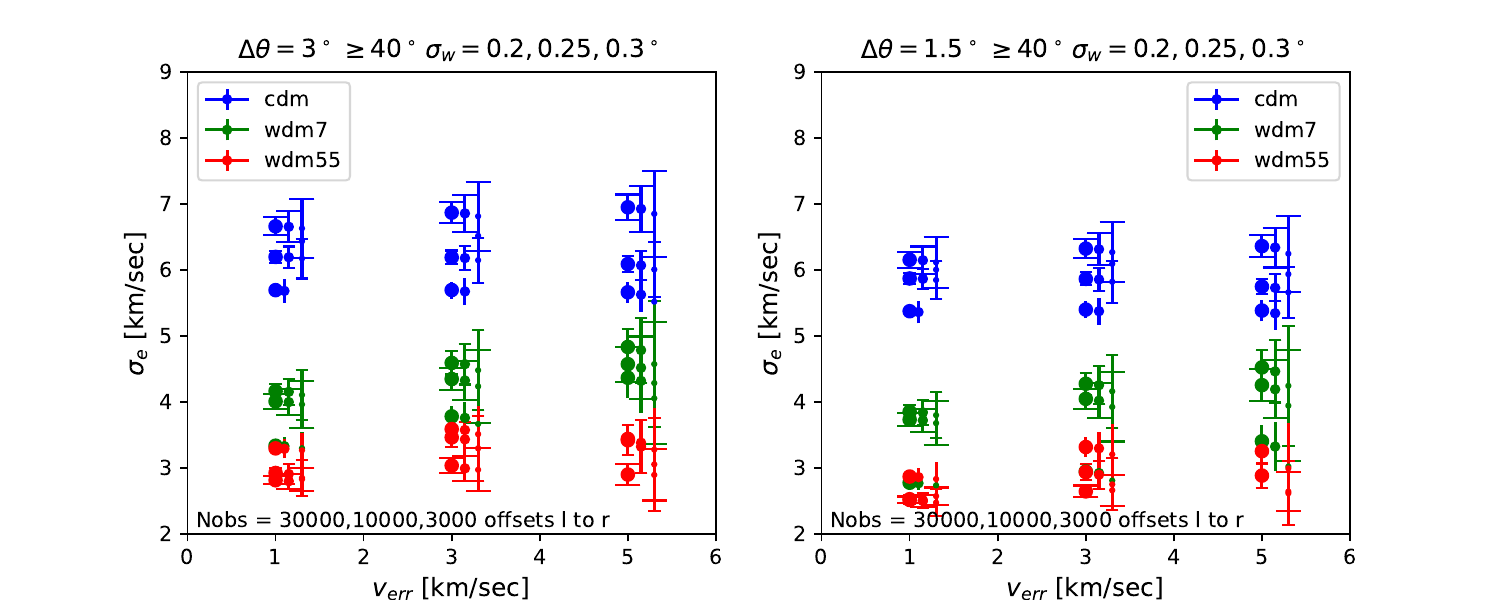}
\includegraphics[angle=0,scale=0.4,trim=50 0 20 0, clip=true]{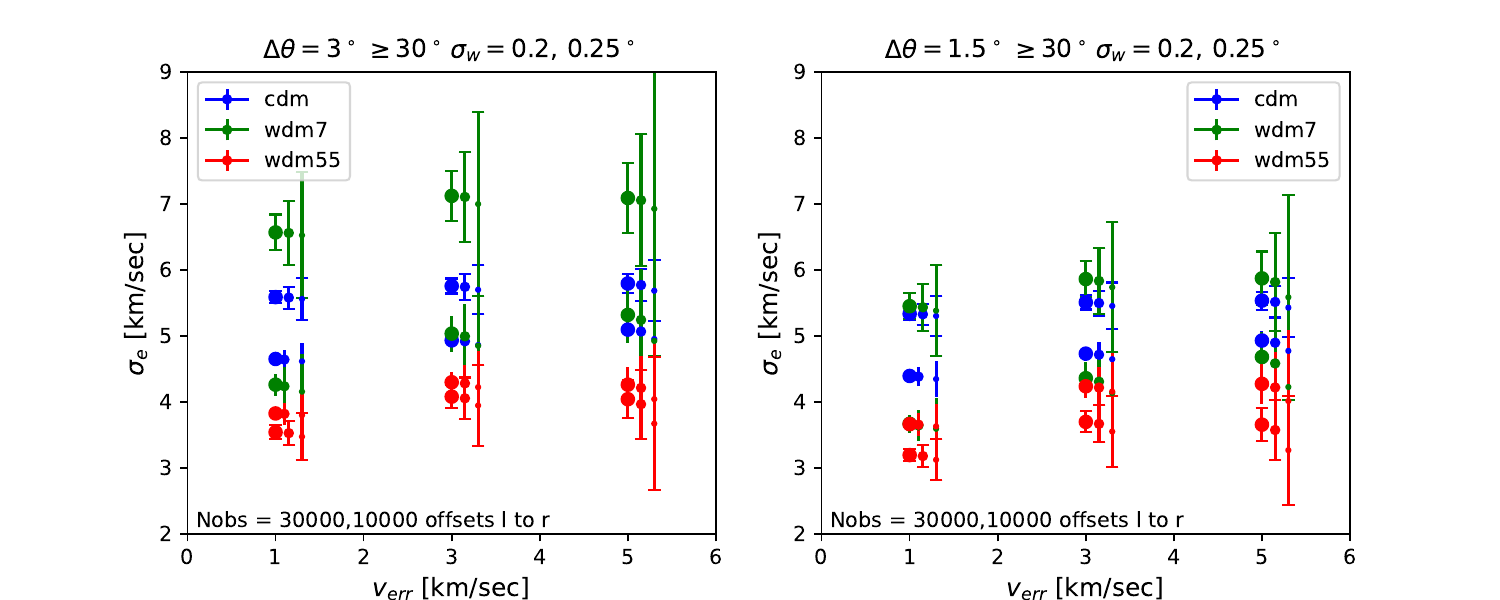}
\includegraphics[angle=0,scale=0.4,trim=50 0 20 0, clip=true]{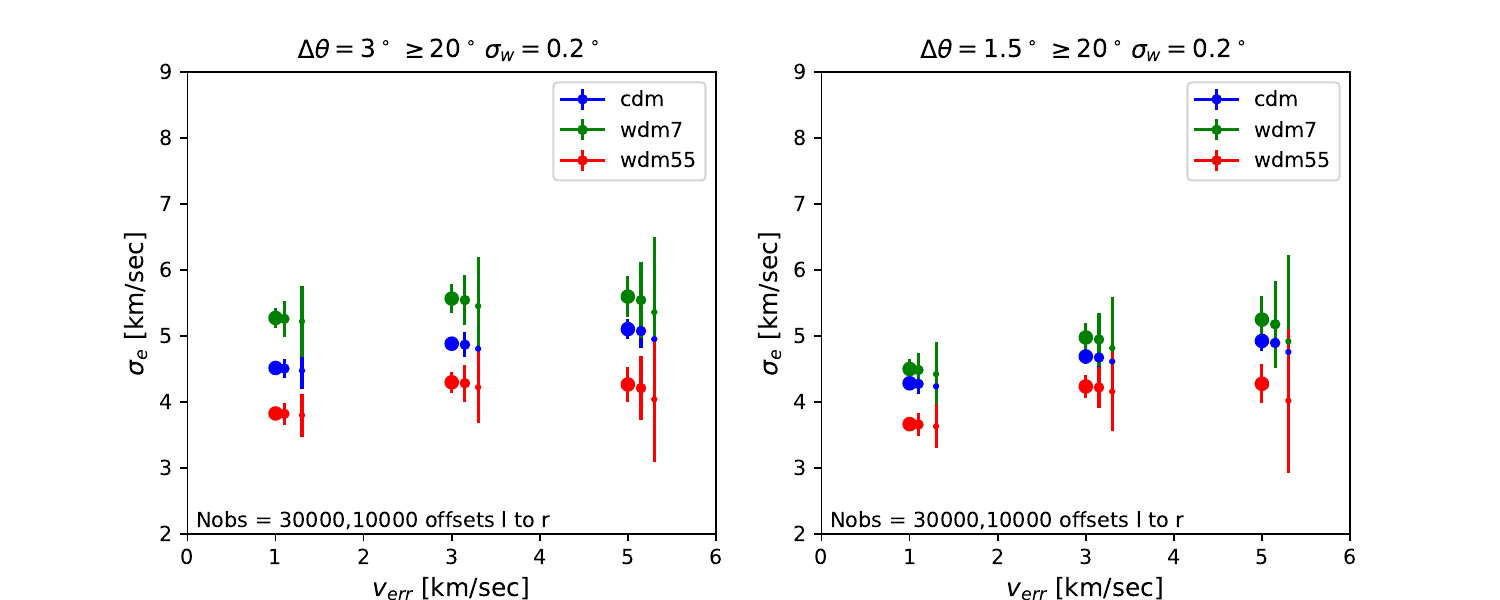}
\end{center}
\caption{
The dependence of $\sigma_e$ and its errors on the velocity errors (x-axis) and the sample size (proportional to point size) for $\geq 40^\circ$ (top ),  $\geq 30^\circ$ (middle) and $\geq 20^\circ$ (bottom row) streams. The length of the caps on the error bars increases with the allowed stream widths indicated in the title. Points are missing when the fit fails to converge. Error bars are 68\% confidence intervals.
}
\label{fig_sigeve}
\end{figure}

Practical measurements of the stream radial velocity distribution function need to consider sample size, allow for velocity errors, and set analysis limits on minimum stream length, maximum allowed stream width, and allow for the angular extent of the data around the stream centerline.   Mock data will guide the design and analysis of any substantial observational program. Here basic aspects of stream selection and sampling are considered as guidelines.  Because the number of low mass subhalos in the galactic halo is reflected in the wings of the radial velocity distribution function the discussion focuses on the exponential term in the  Gaussian plus exponential fit.  

The lower panel of Figure~\ref{fig_paramsge} shows the dependence of $\sigma_e$ on WDM mass and for CDM particles when there are no velocity measurement errors and all available particles are used. A simplified error analysis convolves the velocity profile from the simulations with a Gaussian velocity error and reduces the number of stars in the sample to a target number. The velocity model  for the MCMC fit includes the velocity error as a known input quantity. In the top, left, panel of Figure~\ref{fig_sigeve} the effects of Gaussian velocity errors, increasing from 1 to 5 \kms, and sample size, 3,000, 10,000 and 30,000 particles. The means and errors are shown for  streams  $\geq 40^\circ,\, \sigma_w\leq 0.2-0.3^\circ$ with $\theta \leq \pm 3^\circ$ sampling are shown in Figure~\ref{fig_sigeve}. The error bars are the 68\% (1 sigma) confidence intervals. 

The first result is that reliable results are obtained for velocity errors up to 5 \kms\ , provided that  the velocity error distribution is known. The second result is that for sample sizes of 10,000 stars the $\sigma_e$ error is about 0.3 \kms, meaning that different halo dark matter models can be distinguished at high confidence. A  sample size of 3,000 stars allows models to be distinguished only for velocity errors of 1 \kms. The third result is that streams as short as $20^\circ$ give a weakened result, but only for the thinnest streams,  $\sigma_w\leq 0.2^\circ$. It is notable that the velocity errors do not bias the $\sigma_e$. Including wider streams, $\sigma_w = 0.25^\circ$ and $0.3^\circ$ in the sample increases the number of streams and stars for better averaging. However it also means that streams with more complex velocity structure become part of the sample and the errors tend to increase. 

Reducing the width of the sample region around the stream centerline from $3^\circ$ to $1.5^\circ$ (all angles are measured from the galactic center) decreases the width of the velocity wings about 15\% as shown in the right hand panels of Figure~\ref{fig_sigeve}. The errors on $\sigma_e$ also decrease, so the fractional errors remain approximately the same.  Including shorter streams, down to $\geq 20^\circ$, gives good results for the thin, $\sigma_w\leq 0.2^\circ$ streams. The middle row, streams $\geq 30^\circ$, shows that allowing wider streams causes the velocity wings to increase so much that the results for the 7 keV WDM exceed the CDM values. 

This simplified error analysis finds that the radial velocity distribution function can give useful dark matter particle mass constraints for stream samples up to $0.3^\circ$ width, provided that the streams are $\geq 40^\circ$ long. Sampling to $\pm 1.5^\circ$ from the stream centerline gives smaller $\sigma_e$, but overall the results are comparable to $\pm 3^\circ$ sampling.  Streams as short as $20^\circ$ can be used,  provided $\sigma_w\leq 0.2^\circ$ and only for $\pm 3^\circ$ sampling.  Velocity errors up to 5 \kms\ can be tolerated if the distribution of velocity errors is accurately known.  Velocity errors around 3 \kms\ would allow more leeway in the analysis. A full mock analysis would likely find somewhat larger errors, hence impose stricter requirements on the data to obtain a high quality result.  The difficulties of separating stream stars from the halo field star population are not addressed here. 

\section{Discussion\label{sec_discussion}}

A WDM galactic halo has many fewer dark matter subhalos than a CDM halo. Stellar streams orbiting in WDM halos then have fewer encounters with subhalos than in CDM halos. As a result the width of the random velocity distribution of WDM halo stellar streams is narrower than CDM halo streams, as shown in Figure~\ref{fig_fvge}. The radial velocity distribution is the average along the stream, so relatively simple to construct, not needing the location of a progenitor star cluster.  A simple velocity distribution model, the sum of a  Gaussian and an exponential, fits representative streams. Thinner streams are more sensitive to the numbers of dark halos and longer streams provide better averaging of impacts and more stars to measure.  Streams which provide the most sensitivity to the subhalo numbers are the streams longer than $40^\circ$ with widths up to $\sigma_w= 0.3^\circ$ (as seen from the galactic center)  and streams as short as $20^\circ$ provided $\sigma_w \leq 0.2^\circ$.  Approximately 3-10\% of the streams in the inner 60 kpc of the halo satisfy the long and thin criteria. The stars within  $\pm 1.5$ or $\pm 3^\circ$ of the centerline of the high density part of streams are used to construct the velocity distribution.   The core Gaussian has a width of 1-2 \kms\ with a weak subhalo number dependence.  The width of the exponential wings, $\sigma_e$, increases from 3 \kms\ for a 5.5 keV WDM to 6 \kms\ in CDM, as shown in Figure~\ref{fig_paramsge}. 

The simulations indicate that the number of late time subhalo encounters is only in the range of a few to  tens over the roughly 4~Gyr interval during which stream stars orbit close together.  A consequence is that there is significant variation from stream to stream in the velocity distribution function depending on the details of the subhalo encounter history along its orbit. The higher subhalo densities at earlier times do relatively more stream heating. A representative measurement of the velocity distribution will require a few streams, our results suggest a minimum of three,  to give a robust result. The ability to measure the widths of the radial velocity distribution depends on the velocity errors and the sample size. A simplified analysis, in which the velocity errors are known, finds that velocity errors up to 5 \kms\ can be tolerated provided that there are at least 10,000 stars in the combined streams.  

The width of the stream velocity distribution function is a dynamical consequence of subhalo interactions and orbit diffusion in the aspherical potential of the galaxy.   The number of significant velocity changing subhalo interactions along a stream is relatively low, meaning that at any location along the stream the velocities are usually not well mixed although summing along the entire length of the stream produces a reasonably smooth velocity distribution function.  The stream velocity distribution  effectively measures the average numbers of subhalos in roughly the $10^{6-7} M_\odot$ range that orbit in the same volume as the streams. This subhalo counting approach  naturally complements counting visible dwarf galaxies in dark matter subhalos in the same radial range of the Milky Way, or eventually M31 and other local group galaxies. The new generation of wide field spectrographs have the capability to acquire large number of velocities over a substantial width around a few streams that are  required to make a definitive measurement. 

Simulations like those presented here will be helpful in guiding an observational strategy and interpreting the results. The results here are based on three simulations, all started from the same region cut out from a larger simulation with the small scale power spectrum modified to the WDM versions. The primary halo's mass profile is a good match to the Milky Way but do not contain a Large Magellanic Cloud. The LMC is a relatively large, fast moving potential that distorts the path of streams, but makes little difference to the velocity dispersion within a stream. The simulations show that the stars that are within a few degrees of a stream centerline are largely those that emerged from the progenitor within the last 2-4~Gyr or so. Consequently those stars have little sensitivity to the earlier dynamical history of the galaxy, including the buildup of the disk.  Nevertheless, additional simulations that  sample a range of Milky Way-like assembly histories, the buildup of its baryonic components, and a wider  range of globular cluster origins will be valuable.

\begin{acknowledgements}
This research was supported by NSERC of Canada. This work used the DiRAC@Durham facility managed by the Institute for Computational Cosmology on behalf of the STFC DiRAC HPC Facility (www.dirac.ac.uk). The equipment was funded by BEIS capital funding via STFC capital grants ST/K00042X/1, ST/P002293/1, ST/R002371/1 and ST/S002502/1, Durham University and STFC operations grant ST/R000832/1. DiRAC is part of the National e-Infrastructure. This work used high-performance computing facilities operated by the Center for Informatics and Computation in Astronomy (CICA) at National Tsing Hua University. This equipment was funded by the Ministry of Education of Taiwan, the National Science and Technology Council of Taiwan, and National Tsing Hua University. Computations were performed on the niagara supercomputer at the SciNet HPC Consortium. SciNet is funded by: the Canada Foundation for Innovation; the Government of Ontario; Ontario Research Fund - Research Excellence; and the University of Toronto. CSF acknowledges support by the European Research Council (ERC) through Advanced Investigator grant, DMIDAS (GA 786910). ARJ and CSF acknowledge support from STFC Consolidated Grant ST/X001075/1. APC acknowledges the support of the Taiwan Ministry of Education Yushan Fellowship and Taiwan National Science and Technology Council grant 112-2112-M-007-017-MY3.
\end{acknowledgements}

\software{Gadget4: \citet{Gadget4}, Amiga Halo Finder: \citep{AHF1,AHF2}}, ROCKSTAR: \citep{ROCKSTAR}, NumPy: \citep{numpy}.

Data Availability: Final snapshots, movies, images, and example scripts are at \href{https://www.astro.utoronto.ca/~carlberg/streams}{Streams Data}

\bibliography{Stream}{}

\begin{thebibliography}{}
\expandafter\ifx\csname natexlab\endcsname\relax\def\natexlab#1{#1}\fi
\providecommand{\url}[1]{\href{#1}{#1}}
\providecommand{\dodoi}[1]{doi:~\href{http://doi.org/#1}{\nolinkurl{#1}}}
\providecommand{\doeprint}[1]{\href{http://ascl.net/#1}{\nolinkurl{http://ascl.net/#1}}}
\providecommand{\doarXiv}[1]{\href{https://arxiv.org/abs/#1}{\nolinkurl{https://arxiv.org/abs/#1}}}

\bibitem[{{Angulo} {et~al.}(2013){Angulo}, {Hahn}, \& {Abel}}]{Angulo13}
{Angulo}, R.~E., {Hahn}, O., \& {Abel}, T. 2013, \mnras, 434, 3337,
  \dodoi{10.1093/mnras/stt1246}

\bibitem[{{Behroozi} {et~al.}(2013){Behroozi}, {Wechsler}, \& {Wu}}]{ROCKSTAR}
{Behroozi}, P.~S., {Wechsler}, R.~H., \& {Wu}, H.-Y. 2013, \apj, 762, 109,
  \dodoi{10.1088/0004-637X/762/2/109}

\bibitem[{{Benson} {et~al.}(2013){Benson}, {Farahi}, {Cole}, {Moustakas},
  {Jenkins}, {Lovell}, {Kennedy}, {Helly}, \& {Frenk}}]{Benson13}
{Benson}, A.~J., {Farahi}, A., {Cole}, S., {et~al.} 2013, \mnras, 428, 1774,
  \dodoi{10.1093/mnras/sts159}

\bibitem[{{Binney}(2008)}]{Binney08}
{Binney}, J. 2008, \mnras, 386, L47, \dodoi{10.1111/j.1745-3933.2008.00458.x}

\bibitem[{{Binney} \& {Tremaine}(2008)}]{BT08}
{Binney}, J., \& {Tremaine}, S. 2008, {Galactic Dynamics: Second Edition}
  (Princeton University Press)

\bibitem[{{Bode} {et~al.}(2001){Bode}, {Ostriker}, \& {Turok}}]{Bode01}
{Bode}, P., {Ostriker}, J.~P., \& {Turok}, N. 2001, \apj, 556, 93,
  \dodoi{10.1086/321541}

\bibitem[{{Bond} \& {Szalay}(1983)}]{BS83}
{Bond}, J.~R., \& {Szalay}, A.~S. 1983, \apj, 274, 443, \dodoi{10.1086/161460}

\bibitem[{{Bovy}(2015)}]{galpy}
{Bovy}, J. 2015, \apjs, 216, 29, \dodoi{10.1088/0067-0049/216/2/29}

\bibitem[{{Carlberg}(2012)}]{Carlberg12}
{Carlberg}, R.~G. 2012, \apj, 748, 20, \dodoi{10.1088/0004-637X/748/1/20}

\bibitem[{{Carlberg}(2013)}]{Carlberg13}
---. 2013, \apj, 775, 90, \dodoi{10.1088/0004-637X/775/2/90}

\bibitem[{{Carlberg}(2018)}]{Carlberg18}
---. 2018, \apj, 861, 69, \dodoi{10.3847/1538-4357/aac88a}

\bibitem[{{Carlberg} \& {Agler}(2023)}]{CA23}
{Carlberg}, R.~G., \& {Agler}, H. 2023, \apj, 953, 99,
  \dodoi{10.3847/1538-4357/ace4be}

\bibitem[{{Chandrasekhar}(1942)}]{Chandrasekhar42}
{Chandrasekhar}, S. 1942, {Principles of stellar dynamics}

\bibitem[{{Errani} {et~al.}(2022){Errani}, {Navarro}, {Ibata}, \&
  {Pe{\~n}arrubia}}]{Errani22}
{Errani}, R., {Navarro}, J.~F., {Ibata}, R., \& {Pe{\~n}arrubia}, J. 2022,
  \mnras, 511, 6001, \dodoi{10.1093/mnras/stac476}

\bibitem[{{Foreman-Mackey} {et~al.}(2013){Foreman-Mackey}, {Hogg}, {Lang}, \&
  {Goodman}}]{emcee}
{Foreman-Mackey}, D., {Hogg}, D.~W., {Lang}, D., \& {Goodman}, J. 2013, \pasp,
  125, 306, \dodoi{10.1086/670067}

\bibitem[{{Fukushige} \& {Heggie}(2000)}]{FH00}
{Fukushige}, T., \& {Heggie}, D.~C. 2000, \mnras, 318, 753,
  \dodoi{10.1046/j.1365-8711.2000.03811.x}

\bibitem[{{Gill} {et~al.}(2004){Gill}, {Knebe}, \& {Gibson}}]{AHF1}
{Gill}, S. P.~D., {Knebe}, A., \& {Gibson}, B.~K. 2004, \mnras, 351, 399,
  \dodoi{10.1111/j.1365-2966.2004.07786.x}

\bibitem[{{Grillmair} \& {Dionatos}(2006{\natexlab{a}})}]{GD1}
{Grillmair}, C.~J., \& {Dionatos}, O. 2006{\natexlab{a}}, \apjl, 643, L17,
  \dodoi{10.1086/505111}

\bibitem[{{Grillmair} \& {Dionatos}(2006{\natexlab{b}})}]{Pal5_22deg}
---. 2006{\natexlab{b}}, \apjl, 641, L37, \dodoi{10.1086/503744}

\bibitem[{{Hahn} \& {Abel}(2011)}]{MUSIC}
{Hahn}, O., \& {Abel}, T. 2011, \mnras, 415, 2101,
  \dodoi{10.1111/j.1365-2966.2011.18820.x}

\bibitem[{Harris {et~al.}(2020)Harris, Millman, van~der Walt, Gommers,
  Virtanen, Cournapeau, Wieser, Taylor, Berg, Smith, Kern, Picus, Hoyer, van
  Kerkwijk, Brett, Haldane, del R{\'{i}}o, Wiebe, Peterson,
  G{\'{e}}rard-Marchant, Sheppard, Reddy, Weckesser, Abbasi, Gohlke, \&
  Oliphant}]{numpy}
Harris, C.~R., Millman, K.~J., van~der Walt, S.~J., {et~al.} 2020, Nature, 585,
  357, \dodoi{10.1038/s41586-020-2649-2}

\bibitem[{{Ibata} {et~al.}(2002){Ibata}, {Lewis}, {Irwin}, \&
  {Quinn}}]{Ibata02}
{Ibata}, R.~A., {Lewis}, G.~F., {Irwin}, M.~J., \& {Quinn}, T. 2002, \mnras,
  332, 915, \dodoi{10.1046/j.1365-8711.2002.05358.x}

\bibitem[{{Johnston} {et~al.}(2002){Johnston}, {Spergel}, \&
  {Haydn}}]{Johnston02}
{Johnston}, K.~V., {Spergel}, D.~N., \& {Haydn}, C. 2002, \apj, 570, 656,
  \dodoi{10.1086/339791}

\bibitem[{{Keeley} {et~al.}(2024){Keeley}, {Nierenberg}, {Gilman}, {Gannon},
  {Birrer}, {Treu}, {Benson}, {Du}, {Abazajian}, {Anguita}, {Bennert},
  {Djorgovski}, {Gupta}, {Hoenig}, {Kusenko}, {Lemon}, {Malkan}, {Motta},
  {Moustakas}, {Oh}, {Sluse}, {Stern}, \& {Wechsler}}]{Keeley24}
{Keeley}, R.~E., {Nierenberg}, A.~M., {Gilman}, D., {et~al.} 2024, arXiv
  e-prints, arXiv:2405.01620, \dodoi{10.48550/arXiv.2405.01620}

\bibitem[{{Klypin} {et~al.}(1999){Klypin}, {Kravtsov}, {Valenzuela}, \&
  {Prada}}]{Klypin99}
{Klypin}, A., {Kravtsov}, A.~V., {Valenzuela}, O., \& {Prada}, F. 1999, \apj,
  522, 82, \dodoi{10.1086/307643}

\bibitem[{{Knollmann} \& {Knebe}(2009)}]{AHF2}
{Knollmann}, S.~R., \& {Knebe}, A. 2009, \apjs, 182, 608,
  \dodoi{10.1088/0067-0049/182/2/608}

\bibitem[{{Lovell} {et~al.}(2014){Lovell}, {Frenk}, {Eke}, {Jenkins}, {Gao}, \&
  {Theuns}}]{Lovell14}
{Lovell}, M.~R., {Frenk}, C.~S., {Eke}, V.~R., {et~al.} 2014, \mnras, 439, 300,
  \dodoi{10.1093/mnras/stt2431}

\bibitem[{{Mansfield} {et~al.}(2023){Mansfield}, {Darragh-Ford}, {Wang},
  {Nadler}, \& {Wechsler}}]{Symfind}
{Mansfield}, P., {Darragh-Ford}, E., {Wang}, Y., {Nadler}, E.~O., \&
  {Wechsler}, R.~H. 2023, arXiv e-prints, arXiv:2308.10926,
  \dodoi{10.48550/arXiv.2308.10926}

\bibitem[{{Mao} \& {Schneider}(1998)}]{MS98}
{Mao}, S., \& {Schneider}, P. 1998, \mnras, 295, 587,
  \dodoi{10.1046/j.1365-8711.1998.01319.x}

\bibitem[{{Mateu}(2023)}]{Mateu23}
{Mateu}, C. 2023, \mnras, 520, 5225, \dodoi{10.1093/mnras/stad321}

\bibitem[{{McConnachie}(2012)}]{McConnachie12}
{McConnachie}, A.~W. 2012, \aj, 144, 4, \dodoi{10.1088/0004-6256/144/1/4}

\bibitem[{{Meiron} {et~al.}(2021){Meiron}, {Webb}, {Hong}, {Berczik},
  {Spurzem}, \& {Carlberg}}]{Meiron21}
{Meiron}, Y., {Webb}, J.~J., {Hong}, J., {et~al.} 2021, \mnras, 503, 3000,
  \dodoi{10.1093/mnras/stab649}

\bibitem[{{Miyamoto} \& {Nagai}(1975)}]{MN75}
{Miyamoto}, M., \& {Nagai}, R. 1975, \pasj, 27, 533

\bibitem[{{Moore} {et~al.}(1999){Moore}, {Ghigna}, {Governato}, {Lake},
  {Quinn}, {Stadel}, \& {Tozzi}}]{Moore99}
{Moore}, B., {Ghigna}, S., {Governato}, F., {et~al.} 1999, \apjl, 524, L19,
  \dodoi{10.1086/312287}

\bibitem[{{Nadler} {et~al.}(2021){Nadler}, {Drlica-Wagner}, {Bechtol}, {Mau},
  {Wechsler}, {Gluscevic}, {Boddy}, {Pace}, {Li}, {McNanna}, {Riley},
  {Garc{\'\i}a-Bellido}, {Mao}, {Green}, {Burke}, {Peter}, {Jain}, {Abbott},
  {Aguena}, {Allam}, {Annis}, {Avila}, {Brooks}, {Carrasco Kind}, {Carretero},
  {Costanzi}, {da Costa}, {De Vicente}, {Desai}, {Diehl}, {Doel}, {Everett},
  {Evrard}, {Flaugher}, {Frieman}, {Gerdes}, {Gruen}, {Gruendl}, {Gschwend},
  {Gutierrez}, {Hinton}, {Honscheid}, {Huterer}, {James}, {Krause}, {Kuehn},
  {Kuropatkin}, {Lahav}, {Maia}, {Marshall}, {Menanteau}, {Miquel}, {Palmese},
  {Paz-Chinch{\'o}n}, {Plazas}, {Romer}, {Sanchez}, {Scarpine}, {Serrano},
  {Sevilla-Noarbe}, {Smith}, {Soares-Santos}, {Suchyta}, {Swanson}, {Tarle},
  {Tucker}, {Walker}, {Wester}, \& {DES Collaboration}}]{Nadler21MW}
{Nadler}, E.~O., {Drlica-Wagner}, A., {Bechtol}, K., {et~al.} 2021, \prl, 126,
  091101, \dodoi{10.1103/PhysRevLett.126.091101}

\bibitem[{{Newton} {et~al.}(2021){Newton}, {Leo}, {Cautun}, {Jenkins}, {Frenk},
  {Lovell}, {Helly}, {Benson}, \& {Cole}}]{Newton21}
{Newton}, O., {Leo}, M., {Cautun}, M., {et~al.} 2021, \jcap, 2021, 062,
  \dodoi{10.1088/1475-7516/2021/08/062}

\bibitem[{{Ngan} {et~al.}(2015){Ngan}, {Bozek}, {Carlberg}, {Wyse}, {Szalay},
  \& {Madau}}]{Ngan15}
{Ngan}, W., {Bozek}, B., {Carlberg}, R.~G., {et~al.} 2015, \apj, 803, 75,
  \dodoi{10.1088/0004-637X/803/2/75}

\bibitem[{{Ngan} {et~al.}(2016){Ngan}, {Carlberg}, {Bozek}, {Wyse}, {Szalay},
  \& {Madau}}]{Ngan16}
{Ngan}, W., {Carlberg}, R.~G., {Bozek}, B., {et~al.} 2016, \apj, 818, 194,
  \dodoi{10.3847/0004-637X/818/2/194}

\bibitem[{{Odenkirchen} {et~al.}(2001){Odenkirchen}, {Grebel}, {Rockosi},
  {Dehnen}, {Ibata}, {Rix}, {Stolte}, {Wolf}, {Anderson}, {Bahcall},
  {Brinkmann}, {Csabai}, {Hennessy}, {Hindsley}, {Ivezi{\'c}}, {Lupton},
  {Munn}, {Pier}, {Stoughton}, \& {York}}]{Pal5}
{Odenkirchen}, M., {Grebel}, E.~K., {Rockosi}, C.~M., {et~al.} 2001, \apjl,
  548, L165, \dodoi{10.1086/319095}

\bibitem[{{Onions} {et~al.}(2012){Onions}, {Knebe}, {Pearce}, {Muldrew}, {Lux},
  {Knollmann}, {Ascasibar}, {Behroozi}, {Elahi}, {Han}, {Maciejewski},
  {Merch{\'a}n}, {Neyrinck}, {Ruiz}, {Sgr{\'o}}, {Springel}, \&
  {Tweed}}]{SubHalos12}
{Onions}, J., {Knebe}, A., {Pearce}, F.~R., {et~al.} 2012, \mnras, 423, 1200,
  \dodoi{10.1111/j.1365-2966.2012.20947.x}

\bibitem[{{O'Riordan} \& {Vegetti}(2024)}]{Vegetti24}
{O'Riordan}, C.~M., \& {Vegetti}, S. 2024, \mnras, 528, 1757,
  \dodoi{10.1093/mnras/stae153}

\bibitem[{Shen {et~al.}(2022)Shen, Eadie, Murray, Zaritsky, Speagle, Ting,
  Conroy, Cargile, Johnson, Naidu, \& Han}]{Shen22}
Shen, J., Eadie, G.~M., Murray, N., {et~al.} 2022, The Astrophysical Journal,
  925, 1, \dodoi{10.3847/1538-4357/ac3a7a}

\bibitem[{{Spitzer}(1987)}]{Spitzer87}
{Spitzer}, L. 1987, {Dynamical evolution of globular clusters} (Princeton
  University Press)

\bibitem[{{Springel} {et~al.}(2021){Springel}, {Pakmor}, {Zier}, \&
  {Reinecke}}]{Gadget4}
{Springel}, V., {Pakmor}, R., {Zier}, O., \& {Reinecke}, M. 2021, \mnras, 506,
  2871, \dodoi{10.1093/mnras/stab1855}

\bibitem[{{Springel} {et~al.}(2008){Springel}, {Wang}, {Vogelsberger},
  {Ludlow}, {Jenkins}, {Helmi}, {Navarro}, {Frenk}, \& {White}}]{Springel08}
{Springel}, V., {Wang}, J., {Vogelsberger}, M., {et~al.} 2008, \mnras, 391,
  1685, \dodoi{10.1111/j.1365-2966.2008.14066.x}

\bibitem[{{Vegetti} {et~al.}(2023){Vegetti}, {Birrer}, {Despali}, {Fassnacht},
  {Gilman}, {Hezaveh}, {Perreault Levasseur}, {McKean}, {Powell}, {O'Riordan},
  \& {Vernardos}}]{StrongLens23}
{Vegetti}, S., {Birrer}, S., {Despali}, G., {et~al.} 2023, arXiv e-prints,
  arXiv:2306.11781, \dodoi{10.48550/arXiv.2306.11781}

\bibitem[{{Wang} \& {White}(2007)}]{WW_wdm07}
{Wang}, J., \& {White}, S. D.~M. 2007, \mnras, 380, 93,
  \dodoi{10.1111/j.1365-2966.2007.12053.x}

\bibitem[{{Wetzel} {et~al.}(2016){Wetzel}, {Hopkins}, {Kim},
  {Faucher-Gigu{\`e}re}, {Kere{\v{s}}}, \& {Quataert}}]{Wetzel16}
{Wetzel}, A.~R., {Hopkins}, P.~F., {Kim}, J.-h., {et~al.} 2016, \apjl, 827,
  L23, \dodoi{10.3847/2041-8205/827/2/L23}

\end{thebibliography}
\bibliographystyle{aasjournal}

\end{document}